\documentclass[final]{elsarticle}

\usepackage{lineno, hyperref, pifont, natbib, graphicx}
\renewcommand{\sectionautorefname}{\ifnum\spacefactor>1000 Section\else Section\fi}

\journal{}
\usepackage[margin=1.85cm]{geometry}
\usepackage[USenglish]{babel}
\usepackage{amsmath}
\usepackage{amsfonts}
\usepackage{amssymb}
\usepackage{mathrsfs}
\usepackage{multirow}
\usepackage{cases}
\usepackage{subfig}
\usepackage{booktabs}
\usepackage{hyperref}
\usepackage[dvipsnames]{xcolor}
\usepackage{float}
\usepackage{algorithm} 
\usepackage{comment}
\usepackage{hhline}
\usepackage{algpseudocode}
\usepackage{stmaryrd}
\usepackage{wasysym}

\usepackage{amsthm}
\newtheorem*{definition}{Definition}

\def\equationautorefname~#1\null{%
  Equation~(#1)\null
}

\let\today\relax
\makeatletter
\def\ps@pprintTitle{%
    \let\@oddhead\@empty
    \let\@evenhead\@empty
    \def\@oddfoot{\footnotesize\itshape
         {Submitted preprint} \hfill\today}%
    \let\@evenfoot\@oddfoot
    }
\makeatother

\begin{document}

\begin{frontmatter}

\title{Optimal sizing of 1D vibrating columns accounting for axial compression and self-weight}

\author[dtu]{Federico Ferrari\corref{cor1}}
 \ead{feferr@dtu.dk}


\cortext[cor1]{Corresponding author}
\address[dtu]{Department of Civil and Mechanical Engineering, Technical University of Denmark, Koppels All\'{e}, 2800 Lyngby, Denmark}

\begin{abstract}
We investigate the effect of axial compression on the optimal design of columns, for the maximization of the fundamental vibration frequency. The compression may be due to a force at the columns' tip or to a load distributed along its axis, which may act either independently or simultaneously. We discuss the influence of these contributions on the optimality conditions, and show how the optimal beam design, and the corresponding frequency gain drastically change with the level of compression. We also discuss the indirect effect of frequency optimization on the critical load factors for the tip ($\lambda_{P}$) and distributed ($\lambda_{Q}$) loads. Finally, we provide some quantitative results for the optimal design problem parametrized by the triple ($\lambda_{P}$, $\lambda_{Q}$, $\Omega^{2}$) of buckling and dynamic eigenvalues.
\end{abstract}

\end{frontmatter}

\section{Introduction}
 \label{Sec:intro}

Sizing of beams is a classical topic within optimal structural design, and has been one of the driving applications in the early developments of the field\cite{book:banichuk83, book:rozvani_89a, book:haftka2012}. Despite the simplicity of beam models, closed-form solutions to their optimal design are achievable only in a few cases, involving compliance, volume, or displacement objective or constraints, and linear elastic beams\cite{gjielsvik_71a_minWeightContinuousBeams, huang_71a_optimalDesignBeamsDeflection, rozvany-etal_88a_optimalDesignElasticBeams, pedersen-pedersen_09a_analyticalOptimalDesignBeams}, plastic design\cite{book:rozvani_89a, charret-rozvany_72a_extensionPragerShieldTheory, rozvany_89a_optimalPlasticDesignBeams}, or design against buckling of uniformly compressed, statically determined columns\cite{keller_60a_shapeOfStrongestColumn, rammerstorfer_74a_prestressVibrationBeamOptimization}. In more advanced instances, such as optimization of vibration frequencies, or buckling loads for general boundary conditions and loads, the beams' optimal size can only be computed by numerical methods. 

Nevertheless, important insights about the qualitative behavior of the solution can still be obtained through local asymptotic analysis\cite{book:Hinch91}, and bounds on the optimal frequencies or buckling loads can be given by isoperimetric inequalities\cite{payne_67a_IsoperimetricInequalitiesApplications, tadjbakhsh-keller_62a_strongestColumnIsoperimetric}.
For instance, Niordson\cite{niordson_65a_optimalDesignVibratingBeam} studied the optimal design of a simply supported beam for maximum fundamental frequency of vibration, showing an optimal improvement of $6.6\%$. The analysis was extended to the cantilever beam by Karihaloo and Niordson\cite{karihaloo-niordson_73a_optimumDesignVibratingColumn}, giving a much higher frequency improvement: $425$-$678\%$, depending on the relationship between stiffness and inertia, and then to the optimization of higher frequencies by Olhoff\cite{olhoff_76a_optimizationVibrationsHighOrderFrequencies}. These studies highlighted some delicate matters, which have been the focus of many later works, both dealing with vibration and buckling optimization.

First, at points where the beam's internal actions vanish, the optimal beam size will also go to zero, causing a singularity in one of the generalized displacement or strain components\cite{olhoff-niordson_79a_optimalBeamsColumnsSingularities}. In particular, at locations where only the bending moment vanishes a hinge-like point appears (Type I singularity\cite{gjielsvik_71a_minWeightContinuousBeams}), whereas at locations where both the bending moment and shear force vanish a complete separation of the beam (Type II singularity\cite{olhoff_76a_optimizationVibrationsHighOrderFrequencies}) can occur. This may compromise the well-position of the Sturm-Liouville operator underlying the optimal design problem, \cite{barnes_77a_minmaxProblemsOptimalDesign, barnes_85b_extremalProblemsEigenvalueFunctionals, barnes_88a_extremalProblemsEigenvalueFunctionals}, and several works have focused on finding the correct setting, and admissible class of optimal size distributions to avoid this\cite{krein_55a_onCertainProblems, carmichael_77a_singularOptimalControlVibrationg, gajewsky-piekarski_94a}.

Another delicate point concerns the very existence of non-trivial optimal solutions, depending on the relationship between stiffness and mass, the introduction of design-independent inertia terms, and the particular choice of boundary conditions\cite{prager-taylor_68a_problemsOptimalStructuralDesign}. These matters have been the focus of many mathematical studies, such as those by Brach\cite{brach_68a_extremalFundamentalFrequencyBeam, brach_73a_optimalDesignVibratingStructures}, Holm\aa{}ker \cite{holmaker_87a_optimalDesignVibratingBeam}, and Vepa\cite{vepa73a_existenceSolutionEigenvalueConstraints}.

The literature on optimal beam design against buckling is equally vast \cite{zyczkowski-gajewski_83a_optimalStructuralDesignStability}. Limiting the discussion to recent times, Keller\cite{keller_60a_shapeOfStrongestColumn} started it off by giving a closed-form solution for the optimal design of a simply supported column with maximum buckling load, under a uniform tip compression. Then, Tadjbakhsh and Keller\cite{tadjbakhsh-keller_62a_strongestColumnIsoperimetric} extended to other boundary conditions, including the clamped beam, and this latter case was then reassessed by Olhoff and Rasmussen\cite{olhoff-rasmussen_77a_bimodalOptimumClampedColumn}, accounting for the bimodal character of the optimal solution. The design of a heavy column, where the compression varies along the beam's axis according to self-weight, was addressed by Keller and Niordson\cite{keller-niordson_66a_tallestColumn}, using a combination of asymptotic and numerical methods. In this case, the same technical caveats encountered for frequency optimization are raised, as discussed by Cox and McCarthy\cite{cox-mccarthy_98a_shapeTallestColumn, mccarthy_99a_tallestColumnOptimumRevisited}, or Barnes\cite{barnes_77b_shapeStrongestColumnExtremalEigenvalues, barnes_88a_shapeStrongestColumn}, who introduced minimum requirements on the optimal beam size to ensure well position of the problem. From a more practical stance, the regularization of the problem by introduction of a minimum size had already been proposed by Adali \cite{adali_79a_optimalShapeNonUniformColumn}.

Buckling design under the combined effect of self-weight and a tip, design-independent compression, was studied by Atanackovic, \cite{atanackovic-glavardanov_04a_optimalShapeHeavyColumn}, also considering the bimodal optimization of the clamped column \cite{atanackovic_06a_optimalColumnSelfweight, atanakovic-seyranian_08a_PontryaginOptimalDesign}. Other recent works on the topic are from Sadiku \cite{sadiku_08a_bucklingOptimizationHeavyColumn,sadiku_10a_perturbationOptimizationHeavyColumn}, Seyranian and Privalova\cite{seyranian-privalova_03a_LagrangeProblemOldAndNew}, and Egorov \cite{egorov_03a_onTheLagrangeProblemHollow, egorov_10a_onTheTallestColumn}.

In the author's knowledge, no research work has yet systematically addressed the interplay between buckling and frequency optimization, while considering self-weight. The work by Rammerstorffer\cite{rammerstorfer_74a_prestressVibrationBeamOptimization} is the only one we are aware of discussing the influence of a pre-stress on the beam frequency optimization problem. However, the analysis was limited to uniform pre-stress, simply supported configuration, and a linear control, since the design variable was the Young's modulus.

\begin{figure}[t]
 \centering
  \tabskip=0pt
   \valign{#\cr
  \hbox{%
   \subfloat[]
   {\includegraphics[scale = 1.0, keepaspectratio]{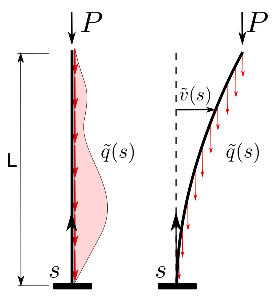}}
 \qquad }\cr
 \noalign{}
  \hbox{%
   \subfloat[SS ($r_{\Omega} = 1.066$)]
   {\includegraphics[scale = 0.35, keepaspectratio]{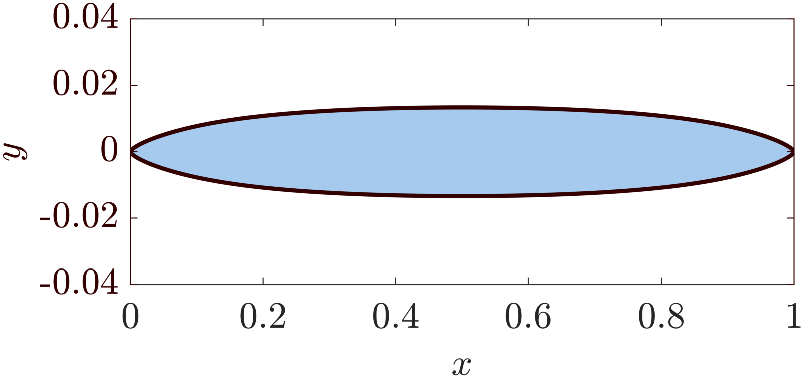}}
  }\vfill
  \hbox{%
   \subfloat[CS ($r_{\Omega} = 1.57$)]
   {\includegraphics[scale = 0.35, keepaspectratio]{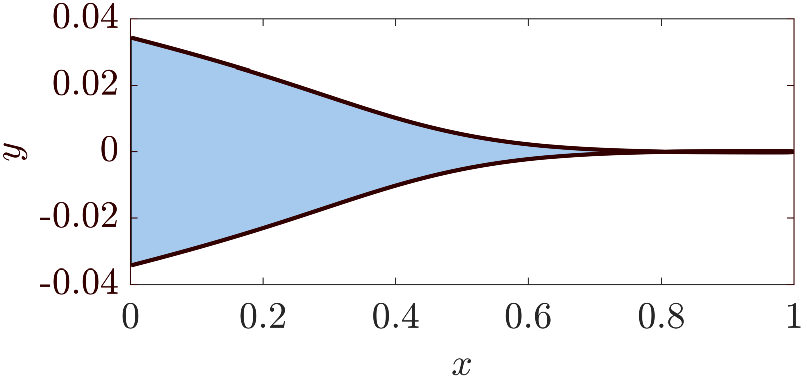}
   \quad}
  }\cr
 \noalign{}
  \hbox{%
  \subfloat[CF ($r_{\Omega} = 6.88$)]
   {\includegraphics[scale = 0.35, keepaspectratio]{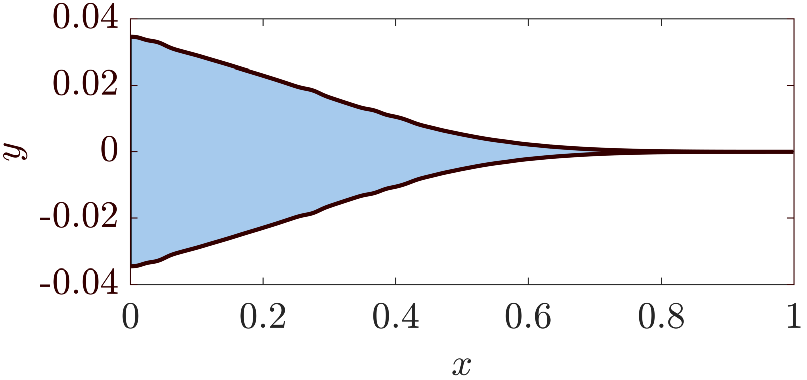}}
  }\vfill
  \hbox{%
   \subfloat[CC ($r_{\Omega} = 4.32$)]
   {\includegraphics[scale = 0.35, keepaspectratio]{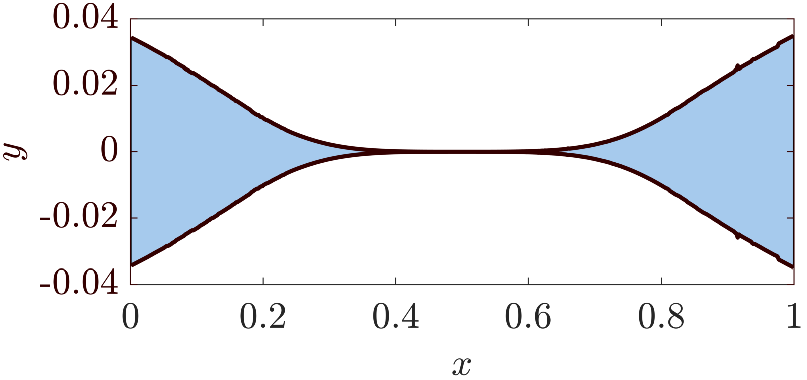}}
  }\cr
  }
 \caption{\fontsize{9}{9} (a) Sketch of a column loaded by the tip force $P$ and distributed axial load $\tilde{q}(s)$, considering the cantilever configuration. (b) Optimal size distribution, maximizing the fundamental frequency of vibration $\Omega_{1}$, for the four cases of boundary conditions: simply supported (SS), cantilever (CF), clamped-pinned (CP), and fully clamped (CC). For each case, $r_{\Omega}$ is the maximum relative frequency gain, given by\cite{niordson_65a_optimalDesignVibratingBeam, olhoff_76a_optimizationVibrationsHighOrderFrequencies}}
 \label{fig:sketchGeometryAndVibrationDesigns}
\end{figure}

As the optimal size of frequency optimized beams shows large variations, and the tendency to vanish at some locations, a question about the beams' stability naturally arises. The designs in \autoref{fig:sketchGeometryAndVibrationDesigns} suggest that the buckling strength of beams optimized purely for vibrations may be severely compromised, to the extent that even self-weight may trigger instability.

Here we aim at providing a broad analysis of the influence of compression on the optimization outcomes. We address the classical problem of maximizing the fundamental frequency of vibration, introducing both constant (size-independent) and variable (size-dependent) compression sources. First, we discuss how these enter the optimally conditions of the problem, also considering the particular case of self-weight. Then, we provide numerical results considering the two compression sources acting either separately or simultaneously. We will show that both substantially change the optimized design, and, generally, also the frequency improvement which can be obtained. Depending on the beam's boundary conditions, and for a compression which is close to the critical one, the natural frequency increases between $7$ to $14.2$ times, when the applied compression is uniform, and up to $15$ times when it is size-dependent. These improvements are much higher than those achieved when optimizing an unloaded beam\cite{niordson_65a_optimalDesignVibratingBeam,karihaloo-niordson_73a_optimumDesignVibratingColumn}, making frequency optimization even more appealing for higher compressed beams\cite{rammerstorfer_74a_prestressVibrationBeamOptimization}.

We also discuss the indirect effect that the optimal design process has on the buckling capacity of the beam. Again, this heavily depends on the amount of compression considered in the optimization, and it is not obvious that the optimized design is stronger than the uniform one as soon as we account for compression in the optimization process.

Finally, we point out that here we consider the technical Bernoulli beam model, following the majority of the research works in the field. Relatively few works have dealt with the Timoshenko beam\cite{huang_71a_effectShearDeformationOptimalDesign, kamat_75a_optimumBeamFrequencyTimoshenko, hua-kentung83a_dynamicOptimizationTimoshenko}, or with geometrically nonlinear models\cite{gierlinsky-mroz_81a_largeDeformationAndShearOptimalBeams, plaut-virgin_11a_elasticaSelfWeight}, or again with non-conservative loads\cite{hanaoka-washizu_80a_optimalDesignBeck, ringertz_94a_optimumBeckColumn}.


\section{Governing equations and statement of the optimization problem}

We consider the initially slender, straight, non-uniform column depicted in \autoref{fig:sketchGeometryAndVibrationDesigns}(a), whose axis occupies the interval $[0,L]\subset \mathbb{R}$. The column is subjected to the tip force of magnitude $P$, and to the distributed axial load with intensity $\tilde{q}(s)$, $s\in (0,L)$, both inducing compression.

Assuming the moderately nonlinear kinematic Bernoulli theory, the Lagrangian density associated with the beam's small amplitude transversal vibrations is
\begin{equation}
 \label{eq:lagrangianDensityAllContributions}
  \ell\left[ s, \tilde{v}( s, t ) \right] = 
   \frac{1}{2}\left[ \rho A \tilde{v}^{2}_{t} +
   \left( P + \int^{L}_{s} \tilde{q}\left( \zeta \right) \: {\rm d}\zeta \right)\tilde{v}^{2}_{s} - EJ \tilde{v}^{2}_{ss} \right]
\end{equation}
where $s$ and $t$ are the space and time coordinates, $\tilde{v}(s,t)$ is the lateral deflection, $A = A(s)$, $\rho = \rho(s)$, $J = J(s)$ and $E = E(s)$ are the beam area, density, modulus of inertia and Young's modulus, respectively. In \autoref{eq:lagrangianDensityAllContributions}, and later on, subscript lowercase letters will denote differentiation with respect to the corresponding variable.

Taking one characteristic dimension of the cross section as variable over the beam length, according to the function $\tilde{y}(s) \in \mathscr{Y} := \left\{ y \in C^{0}(0,L) \, \mid \: \tilde{y}_{\rm min} \leq \tilde{y}(s) \leq \tilde{y}_{\rm max}\right\}$\cite{krein_55a_onCertainProblems, barnes_77b_shapeStrongestColumnExtremalEigenvalues}, the area becomes $A(s) = y_{0} \tilde{y}(s)$, where $y_{0}$ is a constant length \cite{huang_68a, haug-rousselet_80a_DesignSensitivityAnalysis1}. The modulus of inertia can be expressed as
\begin{equation}
 \label{eq:relationshipJp}
 J(s,p) = c_{1} A(s)^{p}
 = \alpha y^{2(2-p)}_{0} \left[ y_{0} \tilde{y}(s)\right]^{p} = \alpha  y^{(4-p)}_{0}\tilde{y}^{p}(s)
\end{equation}
where $c_{1} = \alpha y^{2(2-p)}_{0}$, $\alpha\in\mathbb{R}$, and $p \in \mathbb{N}_{+}$ (usually, $p = \{1, 2, 3\}$). Choosing the $y_{0}$ and $p$, \eqref{eq:relationshipJp} can be particularized to the case of some commonly used cross sections shapes\cite{pedersen-pedersen_09a_analyticalOptimalDesignBeams}.

Introducing the non-dimensional variables $x = s/L$, $\tau = t/T$, $v(x,\tau) = \tilde{v}(s,t)/L$, $y(x) = y_{0}L\tilde{y}(s)/V$, and $q(x) = L\tilde{q}(s)/Q$, where $V$ is the beam's volume and $Q = \int^{L}_{0} \tilde{q}(s) \: ds$ the magnitude of the distributed load, the substitution of \eqref{eq:relationshipJp} in \eqref{eq:lagrangianDensityAllContributions} gives
\begin{equation}
 \label{eq:lagrangianDensityAllContributions-Parametrized-NonDimensional}
  \ell\left[ x, v(x,\tau) \right] = 
   \frac{1}{2}\left[ \varpi y v^{2}_{\tau} + 
   \left( \lambda_{P} + 
   \lambda_{Q}\int^{1}_{x} q(\zeta)
   \: {\rm d}\zeta \right)v^{2}_{x} - 
   y^{p} v^{2}_{xx} \right]
\end{equation}
where we have introduced the non-dimensional parameters
\begin{equation}
 \label{eq:nonDimensionalVariables}
  \lambda_{P} = \frac{P}{V\beta} \, , \ \ 
  \lambda_{Q} = \frac{Q}{V\beta} \, , \ \ 
  \varpi = \frac{\rho L}{T^{2}\beta}
\end{equation}
and $\beta = E\alpha V^{p-1}y^{4-2p}_{0}/L^{p+2}$.

For $\lambda_{P} < \lambda^{({\rm Cr})}_{P}$ and $\lambda_{Q} < \lambda^{({\rm Cr})}_{Q}$, where $\lambda^{({\rm Cr})}_{P}$ and $\lambda^{({\rm Cr})}_{Q}$ are the critical multipliers for the tip force and distributed load, respectively, the beam will undergo small amplitude harmonic oscillations. Therefore, we assume $v(x, \tau) = v(x)e^{-i\omega \tau}$ and $v^{2}_{\tau} = - \omega^{2} v^{2}(x)$ ($\omega > 0$), and imposing stationarity of \eqref{eq:lagrangianDensityAllContributions-Parametrized-NonDimensional} with respect to $\delta v \in \mathscr{V} := H^{2}_{0}(0,1)$ gives the elasto-dynamic linear field equation governing the beam's motion
\begin{equation}
 \label{eq:equilibriumEquation-Parametrized-NonDimensional}
 (y^{p}v_{xx})_{xx} + \left( \lambda_{P} + 
 \lambda_{Q}\int^{1}_{x} q(\zeta) \: {\rm d}\zeta
 \right)v_{xx}
 - \lambda_{Q} q(x) v_{x} - \Omega^{2} y v = 0
 \qquad x \in (0,1)
\end{equation}
where $\Omega^{2} := \varpi\omega^{2}$ is the dynamic eigenvalue, proportional to the square of the angular frequency $\omega$. The boundary terms arising from the integration by parts of \eqref{eq:lagrangianDensityAllContributions-Parametrized-NonDimensional} can be collected in the linear operator
\begin{equation}
 \label{eq:formalBoundaryConditions-Parametrized-NonDimensional}
 B[v,\delta v] := \left.
 \left[ (y^{p}v_{xx})_{x} + \left( \lambda_{P} + 
 \lambda_{Q}\int^{1}_{x} q(\zeta) \: {\rm d}\zeta \right)v_{x}
 \right]\delta v - 
 y^{p} v_{xx}\delta v_{x}
 \right|_{x=\{0,1\}}
\end{equation}
corresponding to one of the following pairs of boundary conditions at $x=\{0,1\}$
\begin{equation}
 \label{eq:casesOfBoundaryConditions-Parametrized-NonDimensional}
 \begin{aligned}
  {\rm either} \quad v = 0 & \qquad 
  {\rm or} \quad (y^{p}v_{xx})_{x} + \left( \lambda_{P} +
  \lambda_{Q}\int^{1}_{x} q(\zeta) \: {\rm d}\zeta \right)v_{x} = 0
  \\
  {\rm either} \quad v_{x} = 0 & \qquad 
  {\rm or} \quad y^{p} v_{xx} = 0
 \end{aligned}
\end{equation}

When the distributed axial load is proportional to self-weight, we have $\tilde{q}(s) = \rho g A(s) = \rho g y_{0}\tilde{y}(s)$ and \eqref{eq:equilibriumEquation-Parametrized-NonDimensional} and \eqref{eq:formalBoundaryConditions-Parametrized-NonDimensional} hold with $q(\zeta) = y(\zeta)$ and $\lambda_{Q} = \rho g \beta^{-1}$.

\subsection{Optimization problem and solution strategy}
 \label{sSec:OptimizationProblem}

We aim at finding the optimal size distribution $y_{\ast} \in \mathscr{Y}$ maximizing the beam's fundamental frequency $\Omega_{1} = \Omega_{1}(\lambda_{P},\lambda_{Q})$, for a prescribed total volume fraction $V$ and compression level, given by the pair $(\lambda_{P},\lambda_{Q})$. This can be stated as
\begin{equation}
 \label{eq:problemMaxEigOmega}
  \begin{aligned}
   \max\limits_{y\in \mathscr{Y}}
   \Omega^{2}_{1}(\lambda_{P}, \lambda_{Q}) := & \left\{
   \min_{v\in\mathscr{V}-0}
   \frac{\langle y^{p}v_{xx}, v_{xx} \rangle_{[0,1]} - \langle (\lambda_{P} + \lambda_{Q} I[x,y])
   v_{x}, v_{x} \rangle_{[0,1]}}
   {\langle y v, v \rangle_{[0,1]}}
   \right\} \\
   {\rm s.t.} & \:
   \langle y, 1 \rangle_{[0,1]} - 1 \leq 0
  \end{aligned}
\end{equation}
where we have introduced the shorthand notation $\langle u,v \rangle_{[a,b]} = \int^{b}_{a} u v \: {\rm d}x$, and set $I[x,y] = \langle q(\xi,y(\xi)), 1 \rangle_{[x,1]}$. In this latter term we account for a general dependence on both $x$ and $y$, allowing for the case of self-weight.

The objective of \eqref{eq:problemMaxEigOmega} is the squared frequency, reduced by both the tip and axially distributed compression, and expressed through the Rayleigh quotient\cite{book:washizu}. This can be related to the one of the unloaded beam $\bar{\Omega}^{2}_{1}$ by
\begin{equation}
 \label{eq:relationshipOmegaOmegaNot}
  \Omega^{2}_{1}(\lambda_{P}, \lambda_{Q})
  = \bar{\Omega}^{2}_{1}
  - \left\langle (\lambda_{P} + \lambda_{Q} I[x,y]) v_{x},
  v_{x} \right\rangle_{[0,1]}
\end{equation}
where, without loss of generality, we assumed $\langle y v, v \rangle_{[0,1]} = 1$. \autoref{eq:relationshipOmegaOmegaNot} expresses the well-known inverse proportionality between the squared natural frequency and axial compression \cite{book:virgin}, and we remark that $\Omega^{2}_{1} = 0$ when either $\lambda_{P} = \lambda^{\rm (Cr)}_{P}$ or $\lambda_{Q} = \lambda^{\rm (Cr)}_{Q}$.

Taking the variation of the Lagrangian associated with \eqref{eq:problemMaxEigOmega}, with respect to $\delta y \in \mathscr{Y}$, we obtain the adjoint equation
\begin{equation}
 \label{eq:AdjEquationAxial-Parametrized-NonDimensional}
  p y^{p-1} v^{2}_{xx} - \lambda_{Q}
  I_{y}[x,y]
  v^{2}_{x} - \Omega^{2}_{1} v^{2}
  - \mu = 0
\end{equation}
where $I_{y}[x,y] = \delta I[x,\delta y]$ and $\mu\in\mathbb{R}_{+}$ is the Lagrange multiplier associated with the volume constraint. Assuming that such constraint is active at the optimum \cite{book:banichuk83}, we have
\begin{equation}
 \label{eq:LagrangeMultiplier-General-NonDimensional}
  \mu_{\ast} = ( p - 1 ) \bar{\Omega}^{2}_{1} + \left\langle 
  \left[\lambda_{P} + \lambda_{Q}
  \left(I[x,y] - yI_{y}[x,y]\right)\right] v_{x}, v_{x} 
  \right\rangle_{[0,1]}
\end{equation}

The optimal beam size distribution is then found by solving \eqref{eq:AdjEquationAxial-Parametrized-NonDimensional} with respect to $y$
\begin{equation}
 \label{eq:OptimalSizeDistribution-General_nonDimensional}
 y_{\ast}(x) = \left[
 \frac{\mu + \lambda_{Q}I_{y}[x,y]v^{2}_{x}+\Omega^{2}_{1}v^{2}}{p v_{xx}}
 \right]^{1-p}
\end{equation}
and once this is known, the optimal displacement $v_{\ast}$ is computed from \eqref{eq:equilibriumEquation-Parametrized-NonDimensional}, paired with the appropriate set of boundary conditions from \eqref{eq:casesOfBoundaryConditions-Parametrized-NonDimensional}, and the eigenvalue $\Omega^{2}_{1\ast}$ is given by the condition $\langle y, 1 \rangle_{[0,1]} = 1$.

The set of nonlinear field equations governing the state, adjoint and optimal size is summarized here
\begin{equation}
 \label{eq:StateAdjointSize_general}
  \begin{aligned}
  0 & = (y^{p}v_{xx})_{xx} + \gamma_{1}v_{xx} - \gamma_{0}v_{x} - \bar{\Omega}^{2}_{1}y v \\
  0 & = p y^{p-1} v^{2}_{xx} - \gamma_{2}v^{2}_{x} - 
  \left(\bar{\Omega}^{2}_{1} - \left\langle \gamma_{1} v_{x}, v_{x} \right\rangle_{[0,1]} \right) v^{2} - 
  (p-1)\bar{\Omega}^{2}_{1} - \left\langle (\gamma_{1} - y\gamma_{2}) v_{x}, v_{x} \right\rangle_{[0,1]} \\
  y & = \left[
  \frac{\gamma_{2}v^{2}_{x} + \left( \bar{\Omega}^{2}_{1} - \left\langle \gamma_{1} v_{x}, v_{x} \right\rangle_{[0,1]} \right) v^{2} +
  (p-1)\bar{\Omega}^{2}_{1} + \left\langle (\gamma_{1} - y\gamma_{2}) v_{x}, v_{x} \right\rangle_{[0,1]}}
  {p v_{xx}}\right]^{1-p}
  \end{aligned}
\end{equation}
where we have set $\gamma_{0} := \lambda_{Q}q(x)$, $\gamma_{1} := \lambda_{P} + \lambda_{Q}I[x,y]$, and $\gamma_{2} := \lambda_{Q}I_{y}[x,y]$.

\autoref{eq:StateAdjointSize_general} is a system of non-linear, coupled, integro-differential equations expressing the optimality conditions for the optimization problem of \eqref{eq:problemMaxEigOmega}. We notice that the design variable $y$ enters non-linearly in both the state and adjoint equations, whereas the state variable $v$ appears only linearly in the state equation, as the coefficients $\gamma_{i}$ do not depend on it. Also, we remark that in \eqref{eq:StateAdjointSize_general} we have written everything in terms of the eigenvalue of the unloaded beam $\bar{\Omega}^{2}_{1}$, and explicitly substituted for the Lagrange multipler expression \eqref{eq:LagrangeMultiplier-General-NonDimensional}, to fully show the influence of each term, in the most general case. However, we will see in the following example that \eqref{eq:StateAdjointSize_general} can be considerably simplified, when tackling specific cases.

\begin{figure}[t]
 \centering
   \subfloat[SS ($l_{1} = l_{2} = 0$, $r_{\Omega}(l_{1}) = 1.066$)]
   {\includegraphics[scale = 0.4, keepaspectratio]{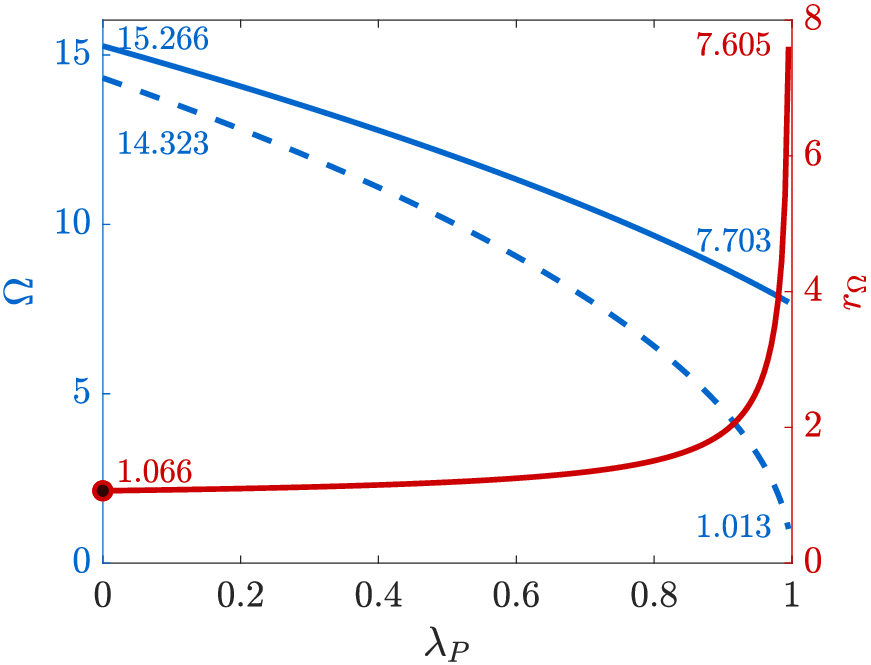}}
   \qquad\qquad
   \subfloat[CF ($l_{1} = 0.56$, $r_{\Omega}(l_{1}) = 2.729$, $l_{2} = 0.975$)]
   {\includegraphics[scale = 0.4, keepaspectratio]{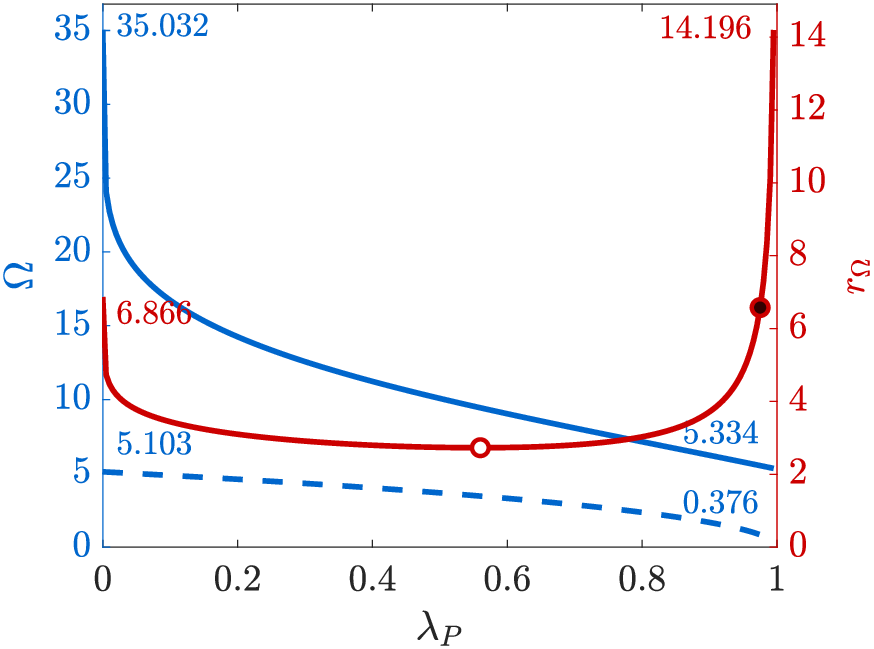}}
   \\
   \subfloat[CS ($l_{1} = 0.3$, $r_{\Omega}(l_{1}) = 1.132$, $l_{2} = 0.755$)]
   {\includegraphics[scale = 0.4, keepaspectratio]{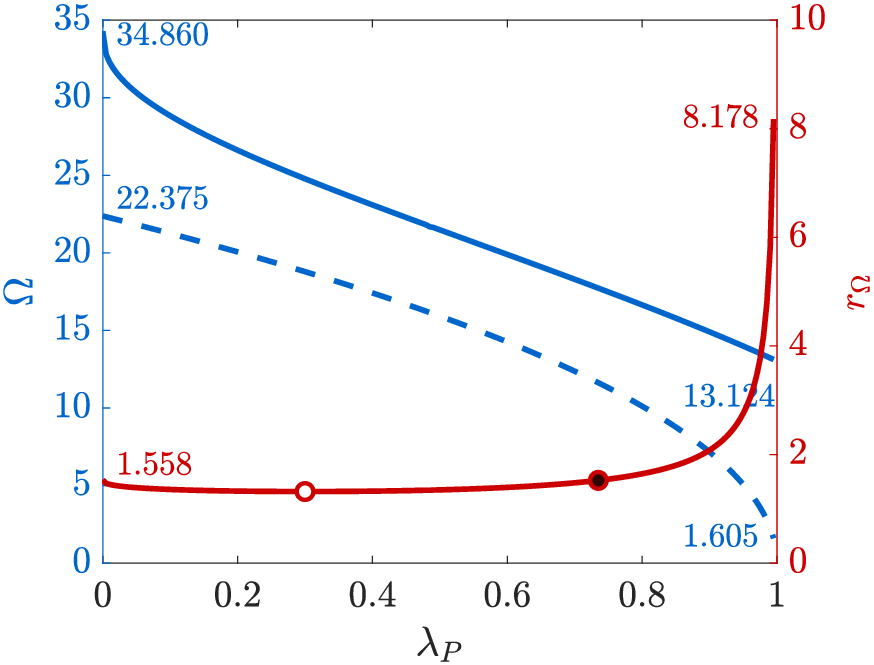}}
   \qquad\qquad
   \subfloat[CC ($l_{1} = 0.525$, $r_{\Omega}(l_{1}) = 1.597$, $l_{2} = 0.975$)]
   {\includegraphics[scale = 0.4, keepaspectratio]{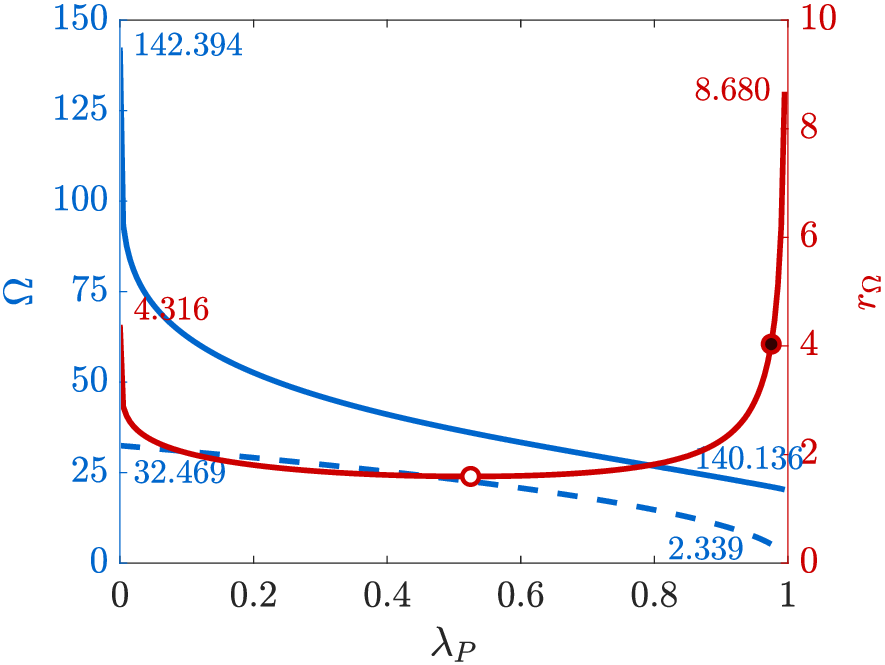}}
 \caption{\fontsize{8}{8} Relationship between the normalized tip compression $\lambda_{P}$ and the frequency of the uniform ($\Omega_{0}$, blue dashed curve), and optimized beams ($\Omega_{\ast}$, blue continuous curve). The frequency gains obtained by solving \eqref{eq:problemMaxEigOmega} for each $\lambda_{P}$, say $r_{\Omega}$, are plotted as a red continuous curve. For the CF, CS, and CC beams $r_{\Omega}$ is non-monotonic and attains the minimum value at $\lambda_{P} = l_{1}$ (white dot), whereas $\lambda_{P} = l_{2}$ (black dot) corresponds to the condition $r_{\Omega}(l_{2}) = r_{\Omega}(0)$}
 \label{fig:numericalComparisonOmegaCurves}
\end{figure}

\section{Numerical results}
 \label{Sec:numericalResults}

We consider a slender beam with aspect ratio $L/y_{0} = 12$, where $y_{0} = 0.5$ is the constant cross section dimension, and also the upper bound $\tilde{y}_{\rm max}$ for the size distribution. The maximum allowed volume is $V = 0.25 y_{0} L$, and the Young modulus and density are $E = 2.5\cdot 10^{7}$ MPa and $\rho = 2.4$ $kg/m^{3}$. We assume similar cross sections\cite{pedersen-pedersen_09a_analyticalOptimalDesignBeams}, setting $p = 2$ and $\alpha = 1/12$ in \eqref{eq:relationshipJp}. Four cases of boundary conditions, also called beam configurations in the following, are considered: cantilever (CF), simply supported (SS), doubly clamped (CC) and clamped-pinned (CP).

The solution strategy for \eqref{eq:problemMaxEigOmega} is based on iterating between the state and adjoint equations \eqref{eq:StateAdjointSize_general}, until both conditions $\|y_{(k)} - y_{(k-1)}\|_{\infty} \leq 10^{-3}$ and $|\Omega_{(k)} - \Omega_{(k - 1)}| \leq 10^{-4}$ are fulfilled. This is carried out repeatedly, starting by setting the lower bound $\tilde{y}_{\rm min} = 10^{-3}$, then reducing it until $\tilde{y}_{\rm min} = 10^{-10}$. In each of these steps, we use the final size distribution from the previous one as initial guess. We will take that a given design shows a Type I, or Type II singularity\cite{olhoff_76a_optimizationVibrationsHighOrderFrequencies} if its final size distribution $y_{\ast}$ hits $\tilde{y}_{\rm min}$ at an isolated point, or at multiple adjacent points, respectively. The state variable $v_{h}$ is discretized by $C^{1}$ elements \cite{book:hughes87}, ensuring nodal continuity of both $v$ and $v_{x}$, whereas the design variable $y_{h}$ is only continuous at nodes. 

In the following, the tip compression is scaled by $\lambda_{P}/\lambda^{\rm Cr}_{P(0)} \in [0, 1)$, where $\lambda^{\rm Cr}_{P(0)} = 
L^{-2}_{c}\pi^{2}EJ_{0}$ is the critical load of the uniform beam, and $L_{c}$ is the characteristic length depending on the boundary conditions \cite{book:timoshenko}. Likewise, the distributed axial load is scaled by $\lambda_{Q}/\lambda^{\rm Cr}_{Q(0)}\in [0,1)$, where $\lambda^{\rm Cr}_{Q(0)}$ is the critical load corresponding to a constant load distribution. To ease the notation, we will denote these multipliers by the symbols $\lambda_{P}$ and $\lambda_{Q}$. As we focus on the fundamental vibration frequency, we drop the subscript ``$1$'', and we denote the frequencies of the uniform and optimized size distributions by $\Omega_{0}$ and $\Omega_{\ast}$, respectively. Also, we will call ``reduced frequencies'' those reduced by the axial compression through \eqref{eq:relationshipOmegaOmegaNot}.

\begin{figure}[t]
 \centering
   \subfloat[SS]
   {\includegraphics[scale = 0.55, keepaspectratio]{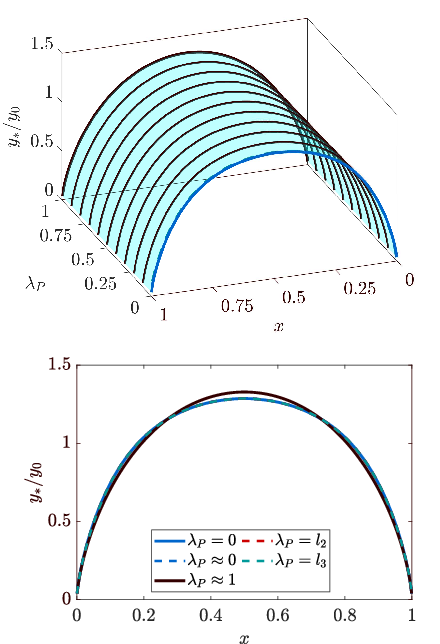}}
   \quad
   \subfloat[CF]
   {\includegraphics[scale = 0.55, keepaspectratio]{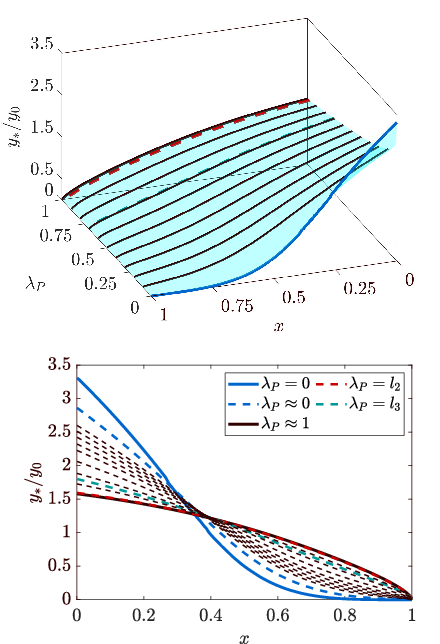}}
   \quad
   \subfloat[CS]
   {\includegraphics[scale = 0.55, keepaspectratio]{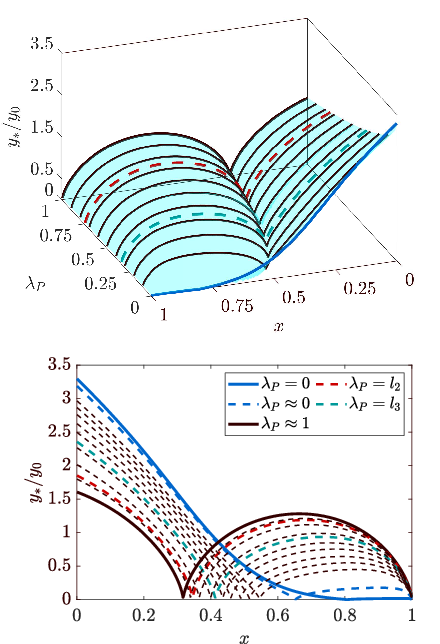}}
   \quad
   \subfloat[CC]
   {\includegraphics[scale = 0.55, keepaspectratio]{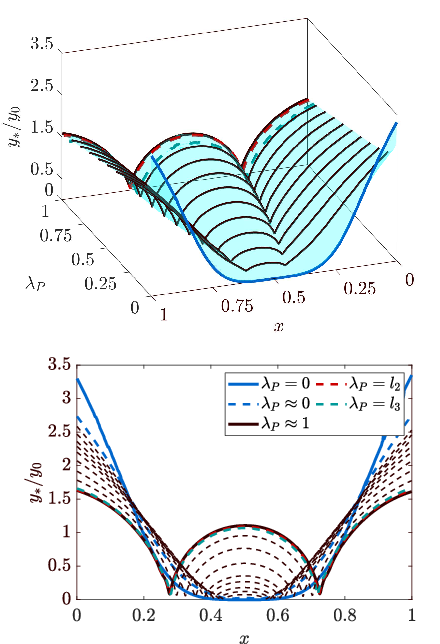}}
 \caption{\fontsize{8}{8} Optimized size distribution $y_{\ast}(x)$, relative to the initial one $y_{0}$, for each beam configuration. On the 3D views, we highlight the designs corresponding to the tip compression values: $\lambda_{P} = 0$ (blue), $\lambda_{P} \approx 1$ (black), and $\lambda_{P} = l_{2}$ (red). This latter corresponds to $r_{\Omega}(l_{2}) = r_{\Omega}(0)$, and the frequency gain equals that of the optimal unloaded beam. The same designs are highlighted on the 2D views, where we also mark the one corresponding to the slightest applied compression $\lambda_{P} \approx 0$ (blue dashed curve)}
 \label{fig:numericalComparisonProfiles3D}
\end{figure}

\subsection{Influence of the tip force alone ($\lambda_{P} \geq 0$, $\lambda_{Q} = 0$)}
 \label{sSec:OnlyLambdaP}

In this case $\gamma_{0} = \gamma_{2} = 0$, $\gamma_{1} = \lambda_{P}$, and the state and adjoint equations in \eqref{eq:StateAdjointSize_general} simplify to
\begin{equation}
 \label{eq:systemOptimalEqs_noQ}
  \begin{aligned}
   0 & = (y^{p}v_{xx})_{xx} + \lambda_{P}v_{xx} - \bar{\Omega}^{2} y v \\
   0 & = p y^{p-1} v^{2}_{xx} - \left(\bar{\Omega}^{2} - \lambda_{P}\left\langle v_{x}, v_{x}\right\rangle_{[0,1]}
   \right) v^{2} - (p-1)\bar{\Omega}^{2} - \lambda_{P}\langle v_{x}, v_{x} \rangle_{[0,1]} \\
 \end{aligned}
\end{equation}
and the design-independent compression influences the solution of \eqref{eq:problemMaxEigOmega} through the state equation and the Lagrange multiplier $\mu$.

The compression, which is constant during the optimization, is applied in the range $\lambda_{P} \in [0, 0.995]$, with $\Delta\lambda_{P} = 1/200$. We refer to the smallest level of non-zero compression ($\lambda_{P} = 0.005$) as $\lambda_{P} \approx 0$, and to the last multiplier as $\lambda_{P} \approx 1$. The BLF of the beam, say $\lambda^{\rm Cr}_{P}$, will be indirectly modified during the optimization, and we must check that $\lambda_{P} \leq \lambda^{\rm Cr}_{P}$ at each step (see \autoref{sSec:indirectEffectBLF}); otherwise the beam will undergo divergence, voiding the assumption of harmonic motion. 

\autoref{fig:numericalComparisonOmegaCurves} shows the influence of the compression level on the frequency of the uniform beam ($\Omega_{0}$, blue dashed line), and of the optimized one ($\Omega_{\ast}$, blue continuous line). While the first follows the well-known (nearly) inverse proportionality between $\Omega^{2}_{0}$ and $\lambda_{P}$ \cite{book:virgin}, the frequency attained in the optimization decreases less than proportionally with the compression level. The relative frequency gain corresponding to each compression level is given by the ratio $r_{\Omega} := \Omega_{\ast}/\Omega_{0}$ (see red line in \autoref{fig:numericalComparisonOmegaCurves}). For $\lambda_{P} = 0$ (optimization of the unloaded beam), $r_{\Omega}$ attains values very close to the theoretical bounds recalled in \autoref{fig:sketchGeometryAndVibrationDesigns} (b). We clearly have $r_{\Omega} \rightarrow \infty$ as $\lambda_{P} \rightarrow 1$; however, we recall that we bound $\lambda_{P}$ to be close but slightly below one. The largest absolute $r_{\Omega}$ is reached by the CF configuration, which for $\lambda_{P} \approx 1$ shows a frequency gain of over 14 times, whereas the SS configuration is the one showing the largest relative improvement between $r_{\Omega}(0)$ (unloaded beam) and $r_{\Omega}(\approx 1)$ (highly compressed beam).

For all the configurations but the SS, the relationship between $\lambda_{P}$ and $r_{\Omega}$ is non-monotonic. For the CF, CS and CC beams, as soon as $\lambda_{P} > 0$ the frequency attained by the optimized beam ($\Omega_{\ast}$) decreases much more rapidly than the frequency of the uniform one, and $r_{\Omega}$ drops. After reaching a minimum at $\lambda_{P} = l_{1}$ (white dot in \autoref{fig:numericalComparisonOmegaCurves}), the frequency gain $r_{\Omega}$ equals the one for the unloaded beam at $\lambda_{P} = l_{2}$ (black dot in \autoref{fig:numericalComparisonOmegaCurves}), then increases further as $\lambda_{P} \rightarrow 1$.

This non-monotonic behaviour can be explained by looking at how the optimized size distribution $y_{\ast}(x)$ changes with the increase of the compression (\autoref{fig:numericalComparisonProfiles3D}). For the SS beam, the optimized size shows minor changes and only gets thicker near the beam center, and more flat near the pinned ends, as $\lambda_{P}$ increases. For all the other beam configurations, the optimized size shows an abrupt change as soon as $\lambda_{P} > 0$, and immediately adapts to sustain the design-independent tip load. The size distribution is prevented from vanishing according to a Type II singularity, whereas we still have points resembling a Type I singularity, where the bending moment vanishes (see \ref{AppA:RemarksOnlLocalBehaviour}). For the CS and CC configurations, these points appear within the beam domain as soon as $\lambda_{P} > 0$ and, for $\lambda_{P} \approx 1$, they are at $x\approx 1/3$ and $x \approx 1/6$, respectively\cite{gjielsvik_71a_minWeightContinuousBeams}. However, we stress that these points do not strictly correspond to singularities, since we always have $y_{\ast} > y_{\rm min}$, as soon as $\lambda_{P} > 0$. 

The optimized design shown as a red dashed line in \autoref{fig:numericalComparisonProfiles3D} corresponds to the compression level $\lambda_{P} = l_{2}$, where we match the frequency gain of the unloaded beam. We also plot, as green dashed line, the one corresponding to $\lambda_{P} = l_{3}$); this is the first designs meeting the same buckling strength of the uniform beam, as will be further discussed in \autoref{sSec:indirectEffectBLF}.

The above discussion shows that for the CF, CS, and CC beams the magnitude of $\lambda_{P}$ has a huge effect on the frequency optimization. Even the slightest design-independent compression prevents the optimal beam size from vanishing according to a Type II singularity, and this drastically lowers the frequency gain. Then, as the compression level increases, the reduction in the initial frequency is so severe that the frequency gain becomes even higher than the one reached when optimizing the unloaded beam. For the SS beam, the frequency gain is always higher for the compressed beam, compared to the unloaded one.

\begin{figure}[t]
 \centering
  \subfloat[\fontsize{7}{7} CF ($r^{\rm (SW)}_{\Omega}(\approx 1) = 19.513$, $l_{4} = 0.385$, $l_{5} = 0.455$)]
  {\includegraphics[scale = 0.375, keepaspectratio]{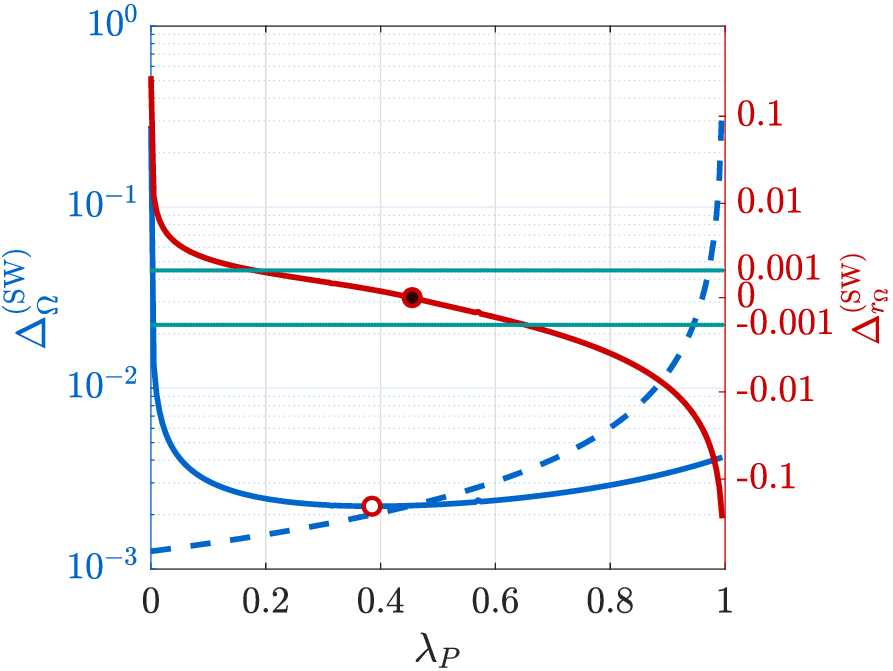}}
  \quad
  \subfloat[\fontsize{7}{7} CS ($r^{\rm (SW)}_{\Omega}(\approx 1) = 8.4650$, $l_{4} = 0.375$, $l_{5} = 0.355$)]
  {\includegraphics[scale = 0.375, keepaspectratio]{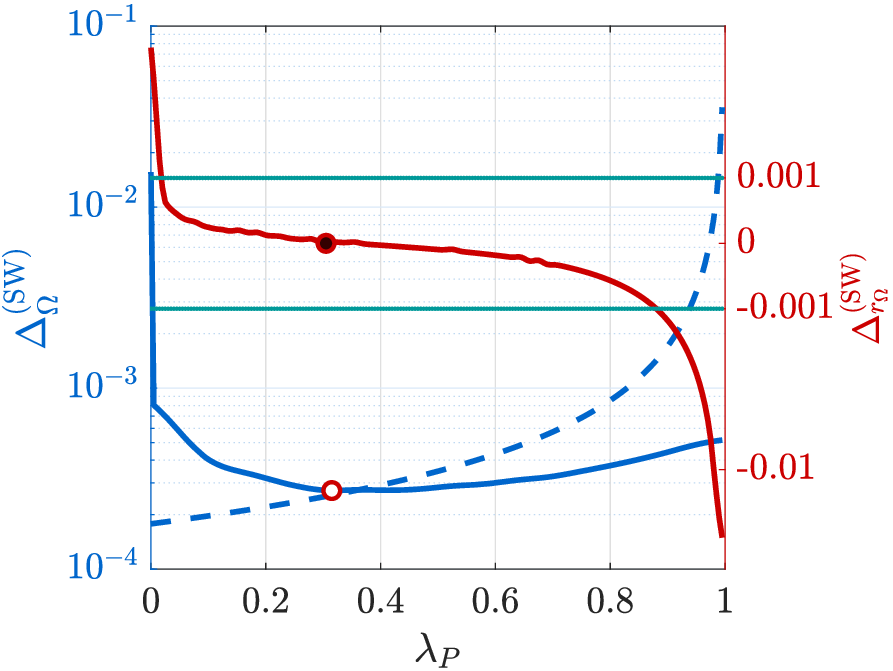}}
  \quad
  \subfloat[\fontsize{7}{7} SS ($r^{\rm (SW)}_{\Omega}(\approx 1) = 8.489$, $l_{4} = 0$)]
  {\includegraphics[scale = 0.375, keepaspectratio]{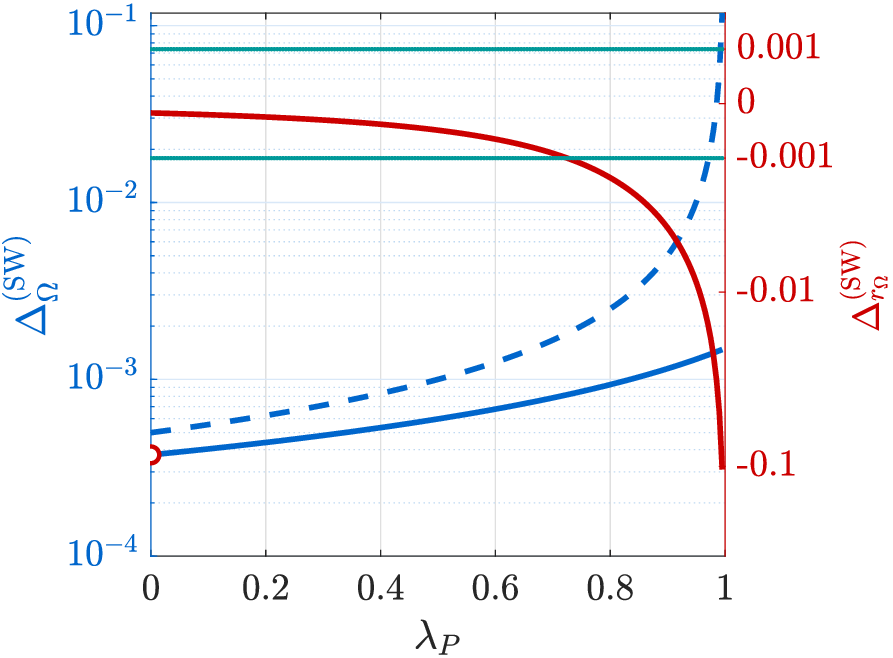}}
 \caption{\fontsize{9}{9} Relationship between the normalized tip compression $\lambda_{P}$ and the relative reduction in the frequency due to the self-weight, for the uniform (blue dashed curve), and optimized (blue continuous curve) beams. The red curve shows the relative reduction of the frequency gains $\Delta^{\rm (SW)}_{r_{\Omega}}$. All curves are scaled according to the $\log$-abs transform. The white dot marks the point $\lambda_{P} = l_{4}$, where we have the minimum of $\Delta^{\rm (SW)}_{\Omega}$, and the black dot marks the point $\lambda_{P} = l_{5}$, where $\Delta^{\rm (SW)}_{r_{\Omega}} = 0$; thus the self-weight has no influence on the frequency gain. The horizontal dotted lines bound the region where $|\Delta^{\rm (SW)}_{r_{\Omega}}| \leq 10^{-3}$. For the CC configuration the trend is basically the same as for the CS, and we refer to \autoref{tab:Table1} for numerical values}
 \label{fig:numericalComparisonInfluenceOfSelfWeight}
\end{figure}

\subsection{Influence of the self-weight ($\lambda_{P} \geq 0$, $q(x) = y(x)$, $\lambda_{Q} = 1$)}
 \label{sSec:selfWeight}

We repeat the analysis of the previous section, now introducing the self-weight contribution. This amounts to the full system of equations \eqref{eq:StateAdjointSize_general}, where $q(x) = y(x)$, $I[x,y] = \left\langle y, 1 \right\rangle_{[x,1]}$, and $I_{y}[x,y] = (1-x)$. Thus, we can simplify $\gamma_{0} = \lambda_{Q}y$, $\gamma_{1} = \lambda_{P} + \lambda_{Q}\left\langle y, 1 \right\rangle_{[x,1]}$, and $\gamma_{2} = \lambda_{Q}(1-x)$.

We remark that, due to the self-weight, buckling may also happen with $\lambda_{P} = 0$. This will be discussed in the next section, and we now fix $\lambda_{Q} = 1$. In the following, we will refer to the beam loaded only by the tip force as the ``light beam'', whereas the ``heavy beam'' also accounts for the self-weight.

We refer to \autoref{fig:numericalComparisonInfluenceOfSelfWeight} where, against the left axis, we plot $\Delta^{\rm (SW)}_{\Omega} := \Omega/\Omega^{\rm (SW)} - 1$, which is the relative difference between the reduced frequency for the light beam ($\Omega$, same as in the previous section), and that of the heavy beam ($\Omega^{\rm (SW)}$). For the uniform design, this quantity is clearly larger than one, as the self-weight introduces an additional compression, further reducing the vibration frequency. Also, it is monotonically increasing with $\lambda_{P}$ for all the beam configurations, and this comes from the coupling of $\lambda_{Q}$ in the $v_{x}$ and $v_{xx}$ terms of the state equation.

Let us focus the discussion on the CF configuration first. For the uniform beam (blue dashed curve), the self-weight alone slightly reduces the frequency ($\Delta^{\rm (SW)}_{\Omega} \approx 0.125\%$ for $\lambda_{P} = 0$). Then, its effect increases with increasing $\lambda_{P}$, and for $\lambda_{P} \approx 1$ we have $\Delta^{\rm (SW)}_{\Omega} \approx 38\%$. The trend is the opposite when looking at the relative difference in the frequency attained by the optimized beam (blue continuous curve). Now the largest difference happens for $\lambda_{P} = 0$ ($\Delta^{\rm (SW)}_{\Omega} \approx 28\%$), as the self-weight is the first source of compression perturbing the optimal design from the one corresponding to the unloaded beam. Then, as $\lambda_{P}$ increases and the size-independent compression dominates the self-weight, $\Delta^{\rm (SW)}_{\Omega}$ quickly drops to a few \permil. However, for the optimized frequency, the ratio $\Delta^{\rm (SW)}_{\Omega}$ is non-monotonous and after reaching the minimum at $\lambda_{P} = l_{4} = 0.385$, the influence of self-weight (slightly) increases again, giving $\Delta^{\rm (SW)}_{\Omega} \approx 0.415\%$ for $\lambda_{P} \approx 1$.

This trend is also reflected by the relative difference between the frequency gains for the light and heavy beams, $\Delta^{\rm (SW)}_{r_{\Omega}} := r_{\Omega}/r^{\rm (SW)}_{\Omega} - 1$ (see red curves in \autoref{fig:numericalComparisonInfluenceOfSelfWeight}). For low magnitudes of the tip compression we have $\Delta^{\rm (SW)}_{r_{\Omega}} > 0$, and the optimization gives a higher frequency gain for the light beam, compared to the heavy one. For $\lambda_{P} = l_{5} \approx 0.455$ we have $\Delta^{\rm (SW)}_{r_{\Omega}} = 0$, and the frequency gains for the heavy and light beams are exactly the same. Then, for higher values of the tip compression we have $\Delta^{\rm (SW)}_{r_{\Omega}} < 0$ and, for $\lambda_{P} \approx 1$ the heavy beam benefits from the optimization about $27.5\%$ more, compared to the light one. However, we observe that apart form the two regions where $\lambda_{P}$ is close to zero or one, the self-weight has a very little influence on the optimization outcomes. For instance, when $0.175\leq \lambda_{P} \leq 0.655$ we have $|\Delta^{\rm (SW)}_{r_{\Omega}}| \leq 10^{-3}$, and therefore the self-weight has a negligible impact on the frequency gain (see horizontal lines in \autoref{fig:numericalComparisonInfluenceOfSelfWeight}).

\begin{figure}[t]
 \centering
   \subfloat[CF]
   {\includegraphics[scale = 0.55, keepaspectratio]{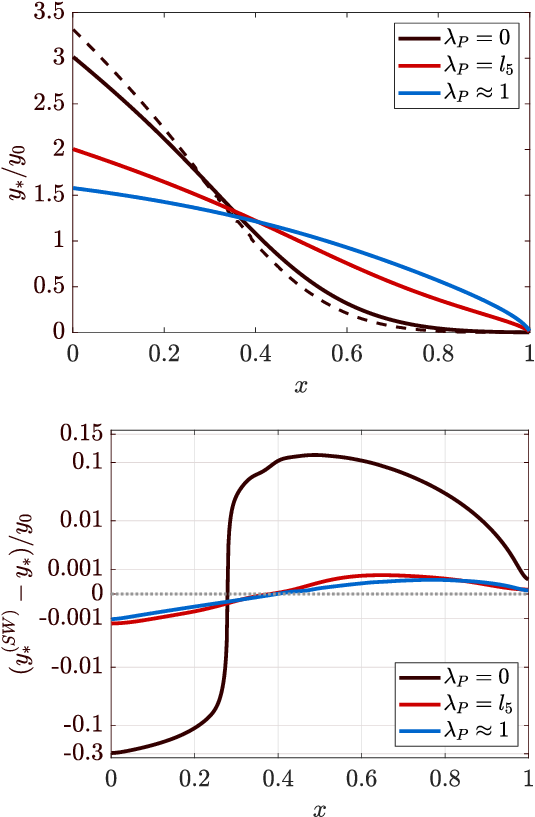}}
   \qquad
   \subfloat[CS]
   {\includegraphics[scale = 0.55, keepaspectratio]{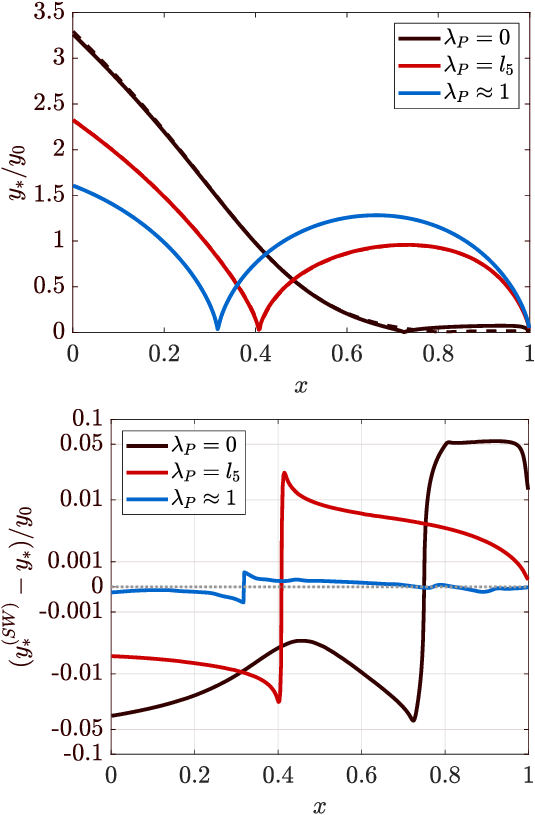}}
   \qquad
   \subfloat[CC]
   {\includegraphics[scale = 0.55, keepaspectratio]{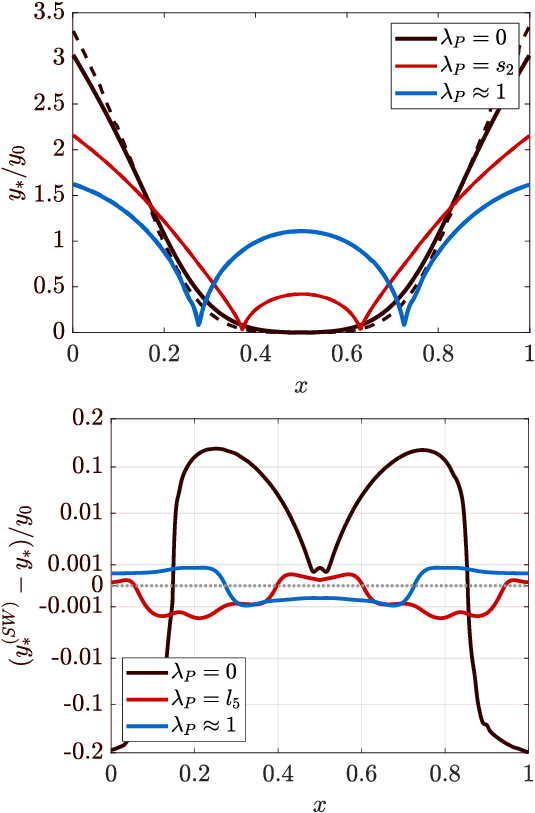}}
 \caption{\fontsize{8}{8} Comparison between the optimized size distribution for the heavy beam (continuous line) and for the light beam (dashed line), for three values of the tip design-independent compression $\lambda_{P}$. We remark that $\lambda_{P} = l_{5}$ correspond to the tip compression at which the self-weight has no influence on the frequency gain. The plots in the bottom row show the log-scaled relative difference between the two size distributions}
 \label{fig:sizeComparison_SW}
\end{figure}

The CS and CC configurations share the same qualitative behaviour of the CF, tough we have different values for the ratios $\Delta^{\rm (SW)}_{\Omega}$, $\Delta^{\rm (SW)}_{r_{\Omega}}$, and location of points $l_{4}$ and $l_{5}$ (see also \autoref{tab:Table1}). Also, both configurations show a much wider range where $|\Delta^{\rm (SW)}_{r_{\Omega}}| \leq 10^{-3}$, and therefore the self-weight has negligible influence on the optimization outcomes ($\lambda_{P}\in [0.01,0.88]$ and $\lambda_{P}\in [0.01,0.86]$ for the CS and CC beams, respectively). This can be clearly seen by the trend of the red curves in \autoref{fig:numericalComparisonInfluenceOfSelfWeight}, for the CF and CS beams. For the SS configuration, both $\Delta^{\rm (SW)}_{\Omega}$ and $\Delta^{\rm (SW)}_{r_{\Omega}}$ are monotonically increasing and decreasing, respectively. Therefore, as $\lambda_{P}$ increases, we gain systematically more when optimizing the heavy beam, compared to the light one.

The effect of self-weight on the optimized size distributions is shown in \autoref{fig:sizeComparison_SW}. Here, the SS beam is not shown, as for this case we have only minor design changes. The plots in the top row compare the relative size distribution of the light (dashed line) and of the heavy beams (continuous line), for three notable tip compression values: $\lambda_{P} = 0$, $\lambda_{P} \approx 1$, and $\lambda_{P} = l_{5}$. For all the beam configurations, we can distinguish between $y^{\rm (SW)}_{\ast}$ and $y_{\ast}$ only for $\lambda_{P} = 0$, as the self-weight has the effect of reducing the size at the clamped ends, while increasing it at the points where the unloaded beam would develop a Type II singularity. For the compression levels $\lambda_{P} = l_{5}$ and $\lambda_{P} \approx 1$, $y^{\rm (SW)}_{\ast}$ and $y_{\ast}$ are almost indistinguishable; however, we can refer to the plots of the bottom row, showing a logarithmic plot of their difference, and recognize a similar trend. For $\lambda_{P} = 0$, the CF beam's sizes show the largest differences due to self-weight. At $x = 0$ the size is reduced of about $30\%$, whereas the maximum increase of about $13.5\%$ occurs at $x \approx 0.5$. For the CS beam, the size's maximum decrease and increase are lower ($4.3\%$ and $5.5\%$), and both occur within the domain, at about $x\approx 0.75$. For the CC beam, the size is reduced of about $24\%$ at the clamped ends, and the maximum increase ($10\%$) occurs, interestingly, at the symmetrical locations $x\approx 0.25$ and $x\approx 0.75$, and not at the mid-point. In all cases, for the higher compression levels the size relative differences quickly drops to very small values. However, it is interesting to note that for the CS configuration we still have size differences up to $\pm 2\%$, corresponding to $\lambda_{P} = l_{5}$, without affecting the frequency gain.

Summarizing, the self-weight contribution sensibly impacts the frequency optimization both when considering zero or high values of the design-independent compression. In particular, frequency optimization becomes highly beneficial for heavy and highly compressed columns. Otherwise, the influence of self-weight on the optimization outcome is very small.

\begin{figure}[t]
 \centering
   \subfloat[]
   {\includegraphics[scale = 0.4, keepaspectratio]{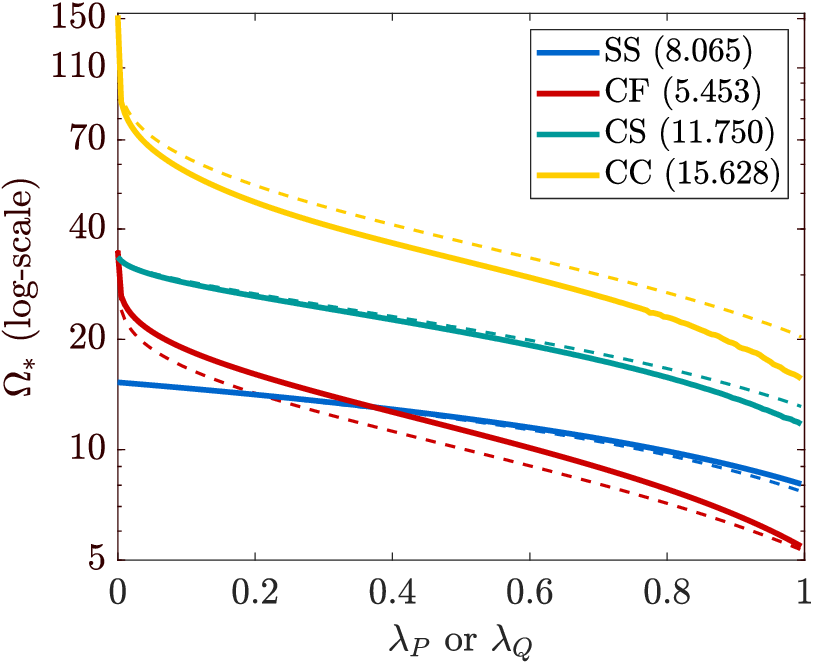}} \qquad
   \subfloat[]
   {\includegraphics[scale = 0.4, keepaspectratio]{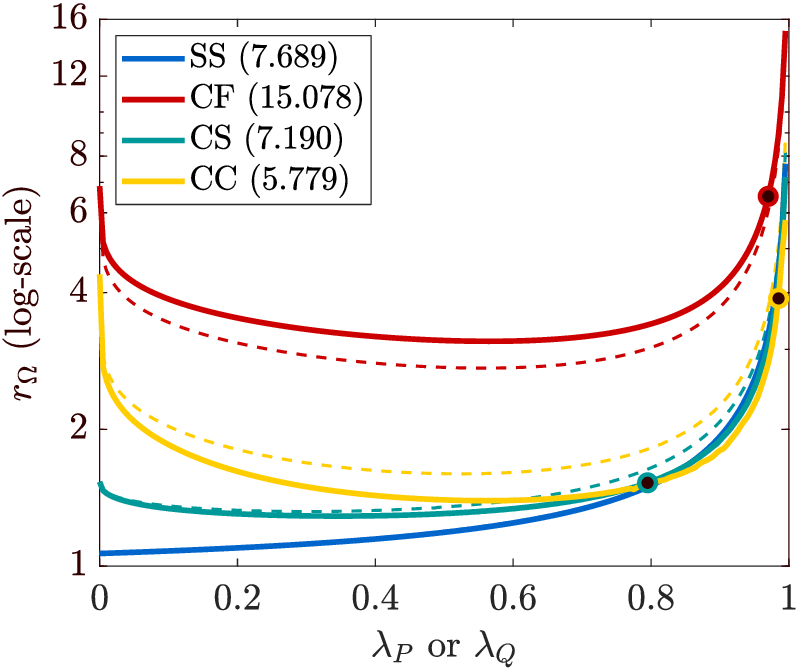}}  \qquad
   \subfloat[]
   {\includegraphics[scale = 0.4, keepaspectratio]{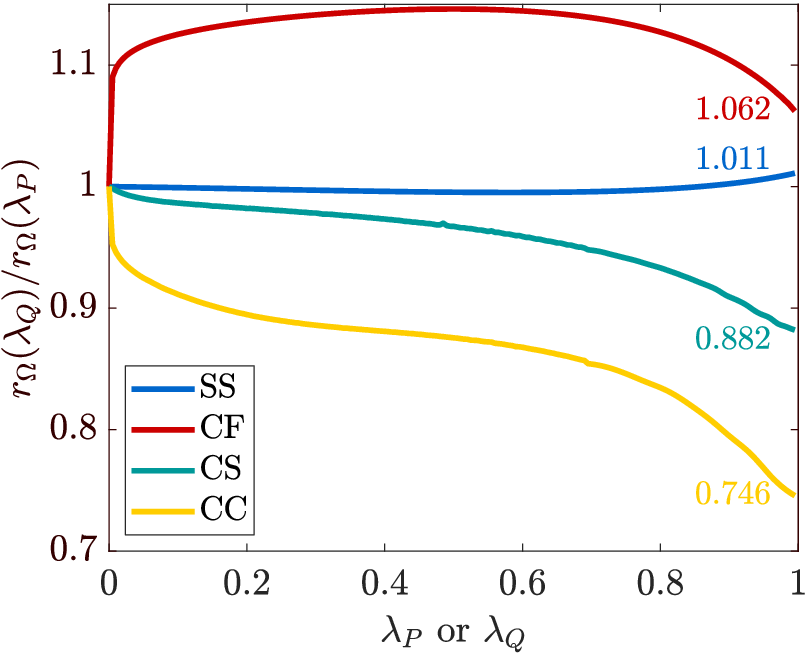}}
 \caption{\fontsize{9}{9} Optimized frequency (a), and the frequency gain (b) for each level of the axial load multiplier $\lambda_{Q}$ and for the four beam configurations. The dashed curves correspond to the case of tip compression $\lambda_{P}$, and are the same as in \autoref{fig:numericalComparisonOmegaCurves}. The black dots in (b) mark the points $\lambda_{Q} =l_{6}$, where $r_{\Omega}(l_{6}) = r_{\Omega}(0)$, where the frequency gain matches that of the unloaded beam. (c) shows the ratio between the frequency gain factors for the cases of distributed axial load and tip compression}
 \label{fig:comparisonFrequenciesLambdaPOnlyLambdaQ}
\end{figure}

\begin{figure}[t]
 \centering
  \includegraphics[scale = 0.45, keepaspectratio]{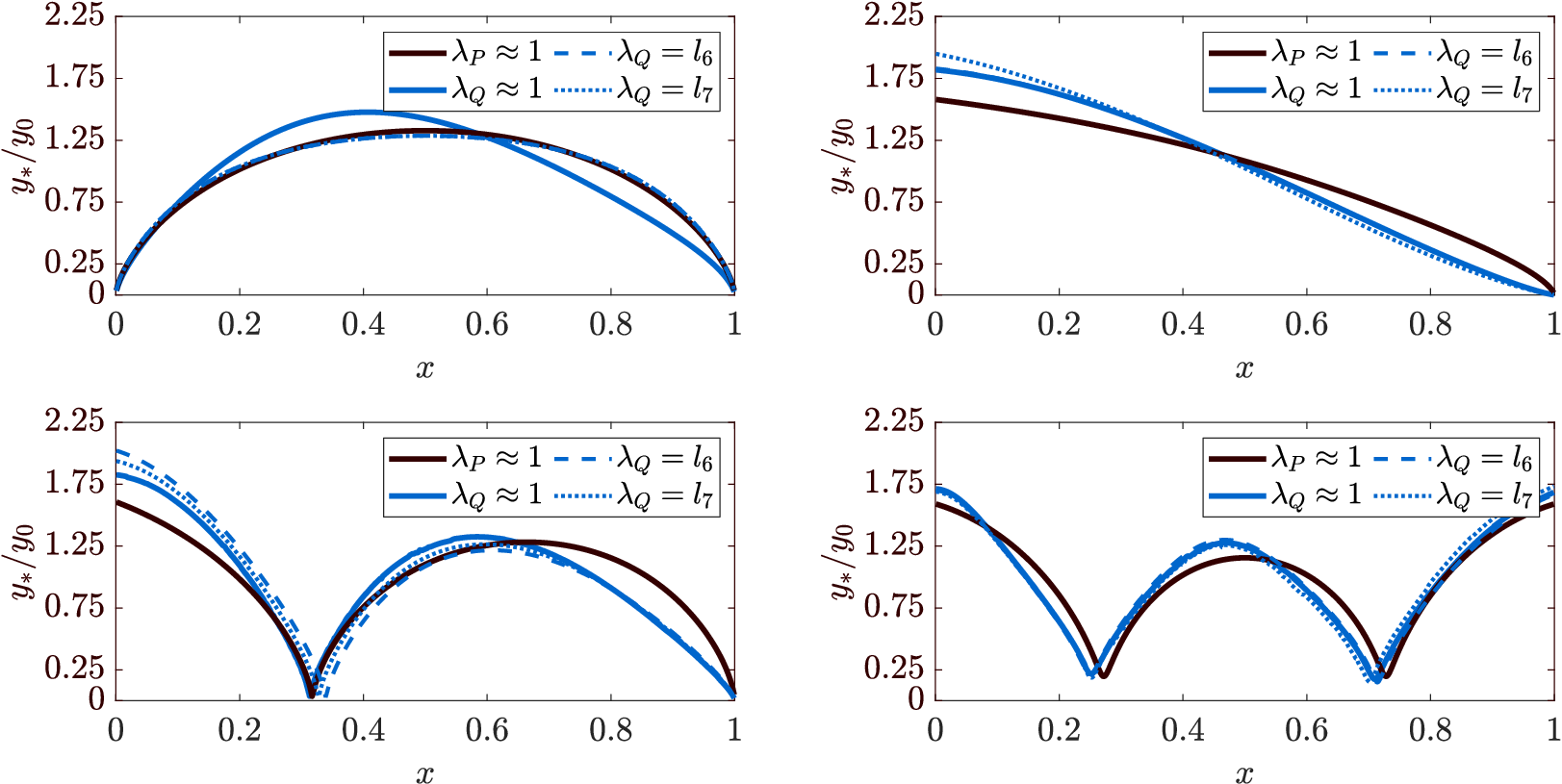}
 \caption{\fontsize{9}{9} Comparison between the optimal size distribution for nearly critical compression, considering the tip force alone ($\lambda_{P}\approx 1$), or the distribued axial load proportional to self-weight alone ($\lambda_{Q} \approx 1$). The two designs shown as a dashed and dotted lines correspond to the compression level $\lambda_{Q} = l_{6}$, where $r_{\Omega}(l_{6}) = r_{\Omega}(0)$, and to $\lambda_{Q} = l_{7}$, respectively. This latter design has the same buckling strength of the uniform one (see \autoref{sSec:indirectEffectBLF})}
 \label{fig:comparisonSizeOnlyLambdaPOnlyLambdaQ}
\end{figure}

\subsection{Influence of the distributed axial load alone ($\lambda_{P} = 0$, $q(x) = y(x)$, $\lambda_{Q} \geq 0$)}
 \label{ssSec:OnlyLambdaQ}

We now remove the size-independent compression, while keeping the distributed axial load proportional to self-weight; thus the coefficients in  \eqref{eq:StateAdjointSize_general} become $\gamma_{0} = \lambda_{Q}y$, $\gamma_{1} = \lambda_{Q}\left\langle y, 1 \right\rangle_{[x,1]}$, and $\gamma_{2} = \lambda_{Q}(1-x)$.

We increase the axial load multiplier up to $\lambda_{Q} = 0.995$, with $\Delta\lambda_{Q} = 1/200$. Again, we have an inverse relationship between $\Omega^{2}_{0}$ and $\lambda_{Q}$ and, for the uniform design, the multiplier causing buckling under self-weight (and thus $\Omega_{0} = 0$) is $\lambda^{\rm Cr}_{Q} = \{942.97, 397.52, 2664.6, 3781.8\}$, for the SS, CF, CS and CC configurations, respectively \cite{book:timoshenko}.

We point out that scaling $\lambda_{Q}$, when considering self-weight, can still be given a meaning as a ``tallest column'' design problem\cite{keller-niordson_66a_tallestColumn, cox-mccarthy_98a_shapeTallestColumn}. Indeed, from \eqref{eq:nonDimensionalVariables} we have $\lambda_{Q} \propto L^{4}$ and $\varpi/\lambda_{Q}\propto L$. Therefore, we may alternatively interpret this as the optimal design of a the tallest column with the same frequency and buckling strength of the initial one. However, the main difference with the classical tallest column problem is that now $\lambda_{Q}$ is set as a parameter.

\autoref{fig:comparisonFrequenciesLambdaPOnlyLambdaQ} (a,b) show the trend of the optimized frequency $\Omega_{\ast} = \Omega_{\ast}(\lambda_{Q})$ and frequency gain $r_{\Omega} = r_{\Omega}(\lambda_{Q})$, for the four beam configurations. For comparison, we also display, as dashed lines, the same curves corresponding to the case of tip compression, identical to those in \autoref{fig:numericalComparisonOmegaCurves}. For all the beam configurations, we highlight the points $\lambda_{Q} = l_{6}$, where $r_{\Omega}(l_{6}) = r_{\Omega}(0)$ (see also \autoref{tab:Table1}). As for the case of tip compression, we have $l_{6} = 0$ for the SS beam, and the frequency gain is monotonically increasing with the load parameter.

For the SS and CF beams, the optimized frequency $\Omega_{\ast}$ corresponding to a given $\lambda_{Q}$ is always larger than that obtained for the same value of $\lambda_{P}$; also, we have $r_{\Omega}(\lambda_{Q}) > r_{\Omega}(\lambda_{P})$. Therefore, for these two configurations frequency optimization leads to larger gains when the beam is loaded by its own self-weight, compared to the tip force compression. The opposite is true for the CS and CC beams, as we now have $\Omega_{\ast}(\lambda_{Q}) < \Omega_{\ast}(\lambda_{P})$ and $r_{\Omega}(\lambda_{Q}) < r_{\Omega}(\lambda_{P})$, for all values of the load multiplier.

These trends are clearly visible from \autoref{fig:comparisonFrequenciesLambdaPOnlyLambdaQ}(c) showing the ratio $r_{\Omega}(\lambda_{Q})/r_{\Omega}(\lambda_{P})$, which is always above one for the SS and CF beams, and always below one for the CS and CC. For the SS configuration the ratio only shows a mild variation, whereas, for the CF beam we have a sudden increase, as the slightest amount of tip compression reduces $\Omega_{\ast}$ much more than the slightest amount of axial load. Then we reach the maximum ratio $r_{\Omega}(\lambda_{Q})/r_{\Omega}(\lambda_{P}) \approx 1.15$ for $\lambda_{Q} = 0.495$, and a slight decrease after. For the CS and CC configurations the decrease of this ratio is monotonic, and very strong for the CC. For the case $\lambda_{Q}\approx 1$ we have frequency gains $1.1\%$ and $6.2\%$ larger, for the SS and CF configurations, and $22\%$ and $36\%$ smaller, for the CS and CC configurations, compared to the case $\lambda_{P}\approx 1$.

\autoref{fig:comparisonSizeOnlyLambdaPOnlyLambdaQ} displays the comparison between the optimized size distribution corresponding to the axially distributed and tip load. The designs are compared for nearly critical compression values ($\lambda_{Q} \approx 1$ and $\lambda_{P} \approx 1$), where the two show the largest differences. We see that the optimal size distributions change from those in \autoref{fig:numericalComparisonProfiles3D}, to a design closer to the tallest column one \cite{cox-mccarthy_98a_shapeTallestColumn}, clearly showing a magnification of the self-weight effect. We also show the designs corresponding to the two notable compression level $\lambda_{Q} = l_{6}$, where the frequency gain matches the one for the unloaded beam, and $\lambda_{Q} = l_{7}$, where the critical load matches that of the initial design (see discussion in the next section).

We point out that, in none of these cases the design develops a Type II singularity and even for $\lambda_{Q} \approx 0$ we have $y_{\ast} = y_{\rm min}$ only at isolated points. This, which may look counter-intuitive, is due to the fact that the multiplier $\lambda_{Q}$ is here a constant parameter, and is not optimized for as in the classical tallest column formulation\cite{keller-niordson_66a_tallestColumn, cox-mccarthy_98a_shapeTallestColumn, mccarthy_99a_tallestColumnOptimumRevisited}.

\begin{figure}[t]
 \centering
   \subfloat[Tip force]
   {\includegraphics[scale = 0.4, keepaspectratio]{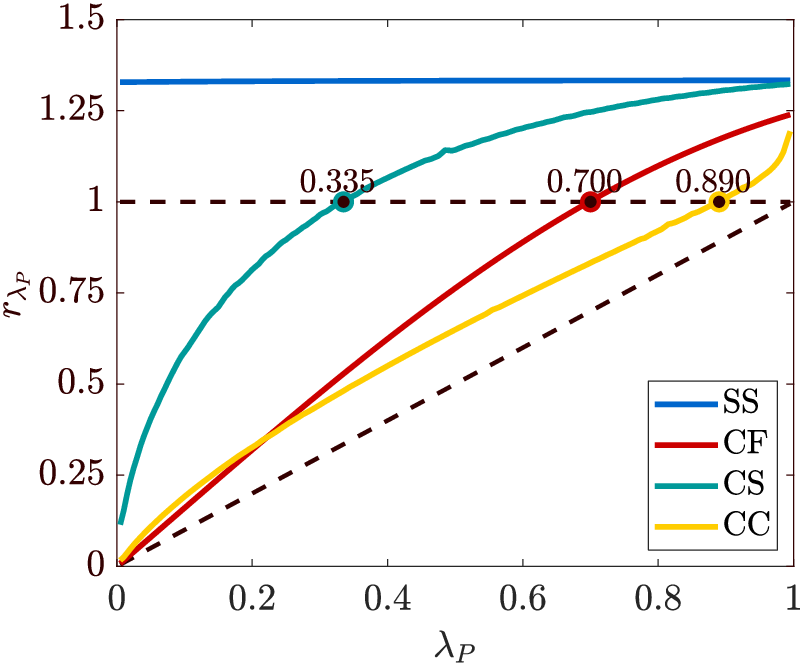}} \qquad\qquad
   \subfloat[Distributed load]
   {\includegraphics[scale = 0.4, keepaspectratio]{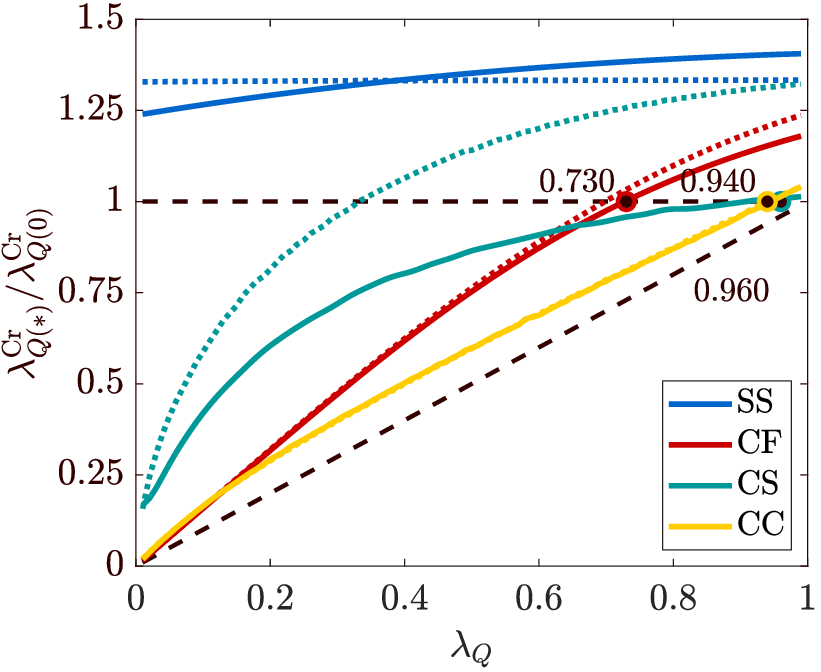}}
 \caption{\fontsize{9}{9} (a) Ratio between the critical load of the optimized and uniform beams, depending on the compression level considered in the optimization \autoref{eq:OptimalSizeDistribution-General_nonDimensional}. The ratios $r_{\lambda_{P}}$ and $r_{\lambda_{Q}}$ are always larger than the current critical load $(\lambda^{\rm Cr})^{-1}$ (dashed line at $45^{\circ}$). The black dots mark the points where $r_{\lambda_{P}} = 1$, and the critical load of the optimized beam equals that of the initial one. The dashed lines in plot (b) report the curves in (a), for comparison}
 \label{fig:numericalComparisonBLFcurves}
\end{figure}

\subsection{Indirect effect on the buckling load factor}
 \label{sSec:indirectEffectBLF}

Finally we investigate the indirect effect of maximizing $\Omega_{1}$ at a given compression level, on the beams' critical load. This latter, which is \emph{not} directly optimized, may either increase or decrease during the optimization. \autoref{fig:numericalComparisonBLFcurves} shows the ratios $r_{\lambda_{P}} := \lambda^{\rm Cr}_{P(\ast)}/\lambda^{\rm Cr}_{P(0)}$ (viz. $r_{\lambda_{Q}} := \lambda^{\rm Cr}_{Q(0)}/\lambda^{\rm Cr}_{Q(\ast)}$), between the critical load of the optimized beam and that of the uniform one, depending on the magnitude of the tip compression $\lambda_{P}$, or axial distributed load $\lambda_{Q}$. The critical load multiplier of the compressed, uniform beam clearly follows the relationship $\lambda^{\rm Cr}_{P(0)} = 1/\lambda_{P}$ (viz., $\lambda^{\rm Cr}_{Q(0)} = 1/\lambda_{Q}$), plotted as a black dashed line in \autoref{fig:numericalComparisonBLFcurves}. Since both $r_{\lambda_{P}}$ and $r_{\lambda_{Q}}$ are always above this line, our assumption of harmonic motion is valid.

For the SS beam, $r_{\lambda_{P}} > 1$ and $r_{\lambda_{Q}} > 1$ for all compression levels, so for this configuration the maximization of the vibration frequency always gives, as a side effect, the increase of the critical load. Also, the range of variation of the two ratios is very narrow for the SS beam, especially when considering the tip compression, for which $1.328 \leq r_{\lambda_{P}} \leq 1.333$. This is expected, as in this case the optimized beam size is only minimally modified (see \autoref{fig:numericalComparisonProfiles3D}(a)). On the other hand, we observe a larger range of variation when considering the axially distributed load, for which $1.23 \leq r_{\lambda_{Q}} \leq 1.406$, and this is consistent with the more marked modification of the beam's profile in this case (see \autoref{fig:comparisonSizeOnlyLambdaPOnlyLambdaQ}(a)).

For the CF, CS, and CC configurations the buckling capacity of the optimized design heavily depends on the compression level considered in the optimization. Up to the values $\lambda_{P} = l_{3}$ and $\lambda_{Q} = l_{6}$, respectively, we have $r_{\lambda_{P}} < 1$ and $r_{\lambda_{Q}} < 1$; therefore the optimized beam will show a lower buckling capacity than the uniform one. Then, for higher compression levels, the optimization also improves the beam's critical load. The designs corresponding to points $l_{3}$ and $l_{6}$, featuring the same critical load as the uniform beam, are marked by green dotted lines in \autoref{fig:numericalComparisonProfiles3D} and \autoref{fig:comparisonSizeOnlyLambdaPOnlyLambdaQ}, respectively.

The first configuration to meet the condition $r_{\lambda_{P}} = 1$, at $l_{3} = 0.335$, is the CS, whereas the last is the CC, at $l_{3} = 0.89$. For the CF configuration, the value of $r_{\lambda_{P}}$ is almost always between those of the CS and CC beams, and we have $l_{3} = 0.7$. We also observe that $r_{\lambda_{P}}$ has an approximately linear trend for the CF and CC configurations, whereas it shows a marked non-linear trend for the CS case. Moreover, for $\lambda_{P} \approx 1$ the CS beam attains a critical load which is slightly below (less than $1\%$) that of the SS beam, whereas the other two beam configurations attain lower values.

When considering the axially distributed load, we see some differences. First, for all beam configurations the same critical load of the uniform design is now reached for higher values of compression (thus $l_{6} > l_{3}$). The trend of $r_{\lambda_{Q}}$ for the CF and CC beams shows little deviations compared to the case of tip compression. On the contrary, the trend for the CS configuration is very different, and this configuration is now the last to reach $r_{\lambda_{Q}} = 1$, for a compression level close to the critial one ($l_{6} = 0.96$).

\section{Conclusions}
 \label{Sec:conclusions}

We have investigated the effect of axial compression on the optimal sizing of beams for maximum fundamental vibration frequency, considering both a tip force generating a constant, control-independent compression, and a distributed axial load proportional to self-weight, giving a control-dependent compression. Some notable numerical results discussed throughout \autoref{Sec:numericalResults} are summarized in \autoref{tab:Table1}.

Both sources of compression have an impact on the frequency gains attainable by the optimization and, for all beam configurations but the SS, also on the optimal size distribution. As soon as compression is introduced, the optimal design deviates from that obtained for the unloaded beam (pure vibrations), and is prevented from developing singularities anywhere in the beam domain. This determines a substantial reduction of the optimal frequency gain, compared to the case of an unloaded beam. However, when looking at high compression levels, frequency optimization becomes much more beneficial, compared to the unloaded case. When the compression is just below the critical one, the frequency gains relative to those for the unloaded beam range between $2$ to $7.1$ for the tip force, and between $11.34$ to $7.2$ times for the distributed load proportional to self-weight.

When paired to the design-independent compression, the self-weight has a very little contribution for most of the ``intermediate'' values of the first. However, it does have an influence when the compression is close to the critical value, where again, its account leads to higher frequency gains, especially for the cantilever and simply supported configuration.

\begin{table}[t]
 \centering
  \begin{footnotesize}
  \begin{tabular}{lc|cccc|ccc|ccc}
   \hline\noalign{\smallskip}
   & \multicolumn{1}{c}{$\lambda_{P}=\lambda_{Q}=0$} 
   & \multicolumn{4}{c}{$\lambda_{P}\geq 0$, $\lambda_{Q}=0$}
   & \multicolumn{3}{c}{$\lambda_{P}\geq 0$, $q(x)=y(x)$, $\lambda_{Q}=1$} 
   & \multicolumn{3}{c}{$\lambda_{P} = 0$, $q(x)=y(x)$, $\lambda_{Q}\geq 0$} \\
   \hline\noalign{\smallskip}
   & $r_{\Omega}$
   & $r_{\Omega}(\lambda_{P}\approx 1)$ & ($l_{1}, r_{\Omega}(l_{1})$) & $l_{2}$ & $l_{3}$
   & $r^{\rm (SW)}_{\Omega}(\lambda_{P}\approx 1)$ & $l_{4}$ & $l_{5}$
   & $r_{\Omega}(\lambda_{Q}\approx 1)$ & $l_{6}$ & $l_{7}$ \\
   \noalign{\smallskip}\hline
   SS & 1.066 &  7.605 & (    0, 1.066) &     0 &     - &  8.489 &     - &     0 &  7.689 &     0 &     - \\ 
   CF & 6.810 & 14.196 & ( 0.56, 2.729) & 0.975 &   0.7 & 19.513 & 0.385 & 0.455 & 15.078 &  0.97 & 0.730 \\
   CS & 1.556 &  8.178 & (  0.3, 1.132) & 0.755 & 0.335 &  8.465 & 0.375 & 0.355 &  7.190 & 0.795 & 0.96 \\
   CC & 4.316 &  8.680 & (0.525, 1.597) & 0.975 &  0.89 &  8.971 & 0.435 &  0.42 &  5.775 & 0.895 & 0.94 \\
   \noalign{\smallskip}\hline
  \end{tabular}
  \end{footnotesize}
  \caption{Summary of the frequency gains and notable points obtained in the numerical examples. The top labels follow the same naming of \autoref{sSec:OnlyLambdaP}-\autoref{sSec:indirectEffectBLF}. $r_{\Omega}(\cdot)$ is the frequency gain, for a given level of compression. The notable compression values are $l_{1}$: $\lambda_{P}$ value that gives the minimum frequency gain, ($l_{2}$ and $l_{6}$): $\lambda_{P}$ and $\lambda_{Q}$ value that gives the same frequency gain as for the unloaded beam, $l_{4}$ minimum frequency gain when considering the self-weight contribution, $l_{5}$, zero influence of the self-weight on the frequency gain, ($l_{3}$ and $l_{7}$) same buckling capacity as the initial beam}
  \label{tab:Table1}
\end{table}

\section*{Acknowledgements}
The author gratefully acknowledges the support of the Villum Foundation, through the Villum Investigator project ``InnoTop''.

\appendix
\section{Remarks on the local behaviour of the dynamic solution}
 \label{AppA:RemarksOnlLocalBehaviour}

Here we summarize the classical local asymptotic analysis for the optimal design of a purely vibrating beam, recalling the definition of the two types of singularities occurring in this case\cite{gjielsvik_71a_minWeightContinuousBeams, karihaloo-niordson_73a_optimumDesignVibratingColumn, olhoff_76a_optimizationVibrationsHighOrderFrequencies}.

For the optimal design of an uncompressed beam ($\lambda_{Q} = \lambda_{P} = 0$), for maximum frequency of vibration, the system governing the optimality \eqref{eq:StateAdjointSize_general} reduces to
\begin{equation}
 \label{eq:systemOptimalEqs_noPnoQ}
  \begin{aligned}
   (y^{p}v_{xx})_{xx} - \bar{\Omega}^{2} y v & = 0 \\
   p y^{p-1} v^{2}_{xx} - \bar{\Omega}^{2} v^{2} - \bar{\Omega}^{2}(p-1) & = 0 \\
   B(y,v) & = 0 \\
   \llbracket M v_{x} \rrbracket_{x_{k}} & = 0  \:, \qquad k = 1, \ldots n_{\rm I}  \\
   \llbracket T v \rrbracket_{x_{k}} & = 0  \: , \qquad k = 1, \ldots n_{\rm II}
 \end{aligned}
\end{equation}
where we have also introduced the boundary operator $B(y, v)$, and we allowed for a finite number of jumps in the optimal energy distribution associated with the bending  moment $M := y^{p}v_{xx}$ and shear force $T := (y^{p}v_{xx})_{x}$. Continuity of these internal actions must hold everywhere within the beam domain, as we do not consider point loads\cite{book:Hartmann1985}, but depending on the vanishing of either or both of them, we can identify two types of points where the optimal size $y_{\ast} \rightarrow 0$, and some of the response components becomes singular.

The first set of points is defined as follows:

\begin{definition}[Type I singularity]
 Let $x^{\rm (I)}_{k} \in [0,1]$ be a point where $M_{\ast}(x^{\rm (I)}_{k}) = 0$, and $T_{\ast}(x^{\rm (I)}_{k}) \neq 0$. Then, $y_{\ast}(\xi) = O(\xi^{1/3})$, where $\xi = |x - x^{\rm (I)}_{k}|$. Also, $v_{\ast}$ is continuous, its derivative $v_{\ast, x}$ has a finite jump, and the curvature $v_{\ast, xx}$ is singular at $x^{\rm (I)}_{k}$.
\end{definition}

At such points, the continuity of the bending moment is automatically fulfilled, $M_{\ast}(x^{\rm I}_{k}) = 0$, and therefore $v_{\ast, x}(x^{\rm (I)}_{k})$ can have a finite jump, provided that $y_{\ast}(x^{\rm I}_{k}) = 0$, such that the energy product $\llbracket y^{p}_{\ast}v^{2}_{\ast, xx} \rrbracket_{x^{\rm I}_{k}}$ remains finite. Mechanically, this point mimics the introduction of a hinge in the beam's domain.

The second set of points leading to a response singularity can be introduced as follows:

\begin{definition}[Type II singularity]
 Let $x^{\rm II}_{k}\in[0,1]$ be a point where $M_{\ast}(x^{\rm II}_{k}) = T_{\ast}(x^{\rm II}_{k}) = 0$. Then, $y_{\ast}(\xi) = O(\xi^{2})$, where $\xi = |x - x^{\rm II}_{k}|$. Also, $v_{\ast}$ has a jump, and both $v_{\ast, x}$ and $v_{\ast, xx}$ are singular at $x^{\rm II}_{k}$.
\end{definition}

At such points both internal actions are vanishing, and this allows for a jump on $v_{\ast}$. Then, the singularity of $v_{\ast,x}(x^{\rm II}_{k})$ and $v_{\ast,xx}(x^{\rm II}_{k})$ requires $y_{\ast}((x^{\rm II}_{k})) = 0$, in order to have finite energy products $\llbracket y^{p}_{\ast}v^{2}_{\ast,xx} \rrbracket_{x^{\rm II}_{k}}$ and $\llbracket (y^{p}_{\ast}v^{2}_{\ast,xx})_{x} v \rrbracket_{x^{\rm II}_{k}}$. Mechanically, this second singularity represents a complete separation of the beam.

\begin{figure}[tb]
 \centering
  \subfloat[Deflection $v_{\ast}$]
  {\includegraphics[scale = .35, keepaspectratio]
  {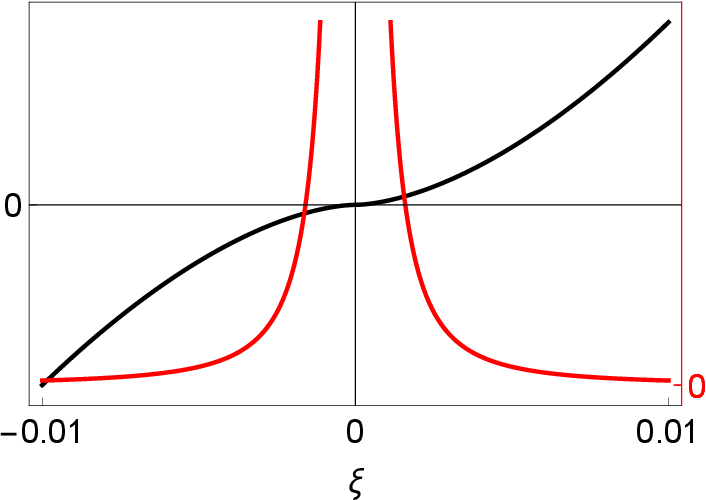}} \qquad
  \subfloat[Rotation $v_{\ast,x}$]
  {\includegraphics[scale = .35, keepaspectratio]
  {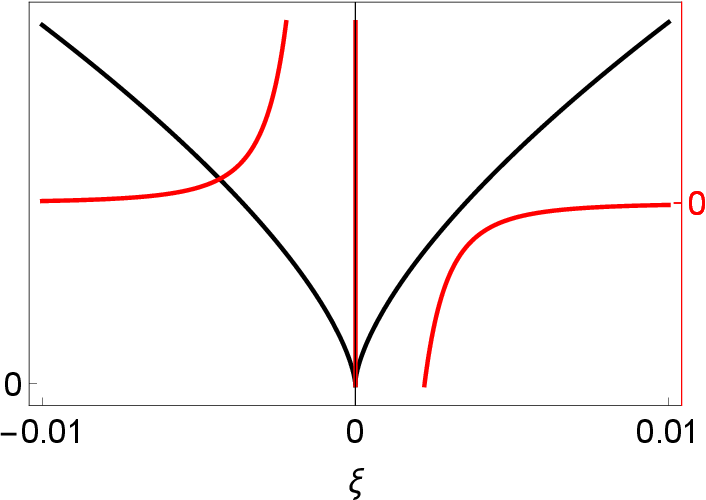}} \qquad
  \subfloat[Curvature $v_{\ast,xx}$]
  {\includegraphics[scale = .35, keepaspectratio]
  {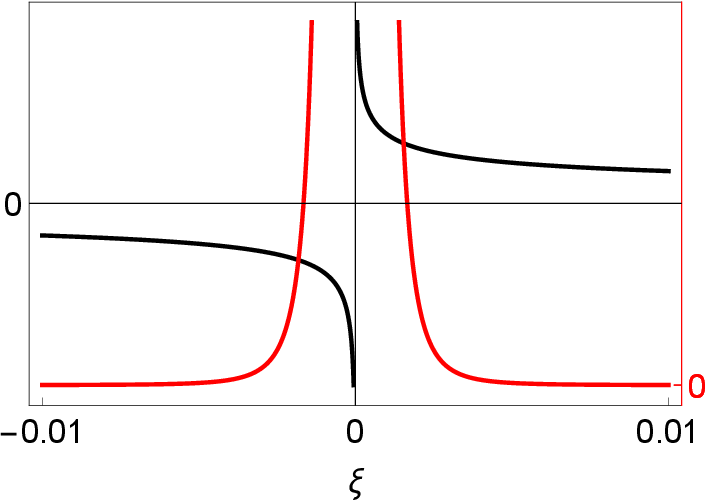}} \\
  \subfloat[Size $y_{\ast}$]
  {\includegraphics[scale = .35, keepaspectratio]
  {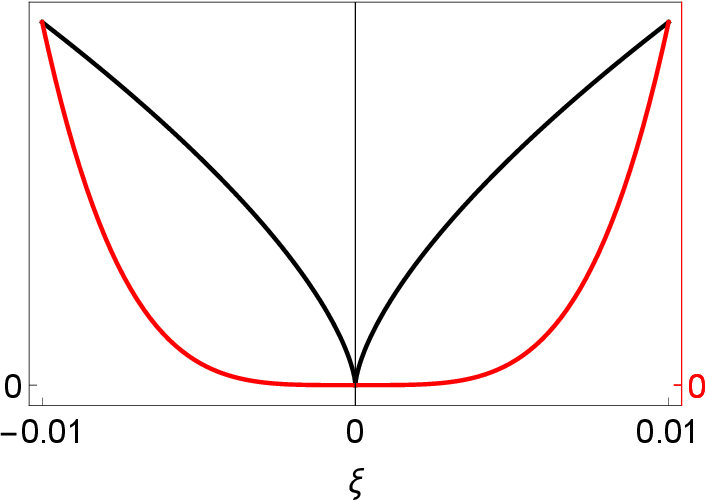}} \qquad
  \subfloat[Bending moment $M_{\ast}$]
  {\includegraphics[scale = .35, keepaspectratio]
  {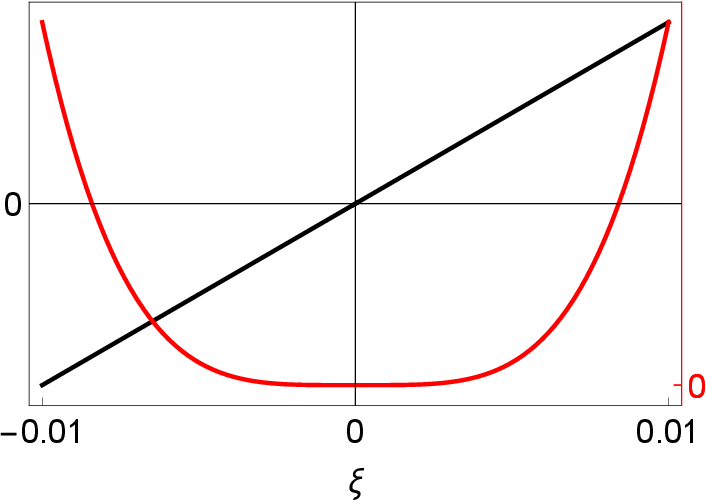}} \qquad
  \subfloat[Shear force $T_{\ast}$]
  {\includegraphics[scale = .35, keepaspectratio]
  {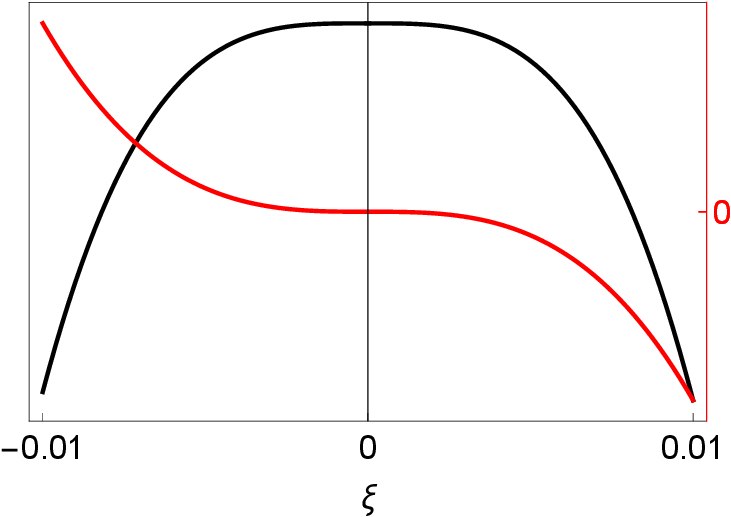}}
 \caption{Optimal response functions, size distribution, and internal actions near a Type I (black curves, plotted against the left axis), and Type II (red curves, plotted against the right axis) point, according to \eqref{eq:App-OrderOfFunctions_TypeI} and \eqref{eq1:SubstitutionRespFunS2}}
  \label{fig:App-BehaviourSingularities}
\end{figure}

The behaviour of the response functions near these two types of points can be deduced by an asymptotic analysis on the following non-linear differential equation
\begin{equation}
 \label{eq:App_OptimalitySystemPureDynamic}
  \left[\left( \frac{\bar{\Omega}^{2}(v^{2}_{x} + p - 1)}
  {p v^{2}_{xx}} \right)^{p(1-p)} v_{xx} \right]_{xx} -
  \left(\frac{\bar{\Omega}^{2}(v^{2} + p - 1)}{p v^{2}_{xx}} \right)^{1-p} v = 0
\end{equation}
obtained by eliminating the expression for $y_{\ast}$ from the state equation in \eqref{eq:systemOptimalEqs_noPnoQ}. Then we represent the deflection $v_{\ast}(\xi)$ according to a regular, or a singular expansion \cite{book:Hinch91}.

In the first case we assume $v^{\rm (I)}(\xi) = a_{0} + a_{1}\xi^{r}$, where $r\in\mathbb{Q}$ and $(a_{0}, a_{1})\in\mathbb{R}$, and substituting within \eqref{eq:App_OptimalitySystemPureDynamic}, we obtain the residual
\begin{equation}
 \label{eq:localExpansionFreeVibDEQ_regular}
 \begin{aligned}
 \mathcal{R}[v^{\rm (I)}] =
 \frac{\bar{\Omega}^{2}\xi^{4-3r}}{t_{0}}
 \left[
 (a^{4}_{0}+3(1+2a^{2}_{0}))t_{1} +
 o(\xi) \right] = 0
 \end{aligned}
\end{equation}
where $t_{0} = 4a_{2}r^{3}(r-1)^{3}$ and $t_{1} = 3(r-2)(3r-5)$. Non-trivial solutions to \eqref{eq:localExpansionFreeVibDEQ_regular} require $t_{1} = 0$, while $t_{0} \neq 0$. This correspond to the pair of solutions $r_{1} = 5/3$ and $r_{2} = 2$, which plugged in \eqref{eq:localExpansionFreeVibDEQ_regular} give
\[
 \lim_{\xi \rightarrow 0^{+}}
 \mathcal{R}\{r_{1},r_{2}\} =
 \{0, \frac{(a_{0}+a^{3}_{0})\bar{\Omega}^{2}}
 {8 a^{2}_{2}}\}
\]
and therefore only the solution $r = 5/3$ is acceptable. The behaviour of the functions near a Type I points is (see \autoref{fig:App-BehaviourSingularities})
\begin{equation}
 \label{eq:App-OrderOfFunctions_TypeI}
  \begin{split}
       v_{\ast}\left( \xi \right) & = a_{0} + a_{1}\xi + \xi^{ 5/3} \: , \quad
   v_{\ast, x}\left( \xi \right)   = a_{1} + a_{2} \xi^{ 2/3} \: , \quad
  v_{\ast, xx}\left( \xi \right)   = a_{3} \xi^{-1/3} \\
       y_{\ast}\left( \xi \right) & = a_{4} \xi^{1/3} + O(\xi)  \: , \quad
       M_{\ast}\left( \xi \right)   = \xi + O(\xi^{2})        \: , \quad
       T_{\ast}\left( \xi \right)   = a_{5} + O(\xi)
  \end{split}
\end{equation}
where the expression of the $a_{i}$ coefficients can be found in \cite{olhoff_76a_optimizationVibrationsHighOrderFrequencies}.

If we consider the singular expansion $v^{\rm (II)}(\xi) = b_{0} \xi^{-s}$, $s\in \mathbb{N}_{+}$, $b_{0}\in\mathbb{R}$ the residual of \eqref{eq:App_OptimalitySystemPureDynamic} becomes
\begin{equation}
 \label{eq:localExpansionFreeVibDEQ_singular}
 \begin{aligned}
  \mathcal{R}[v^{\rm (II)}] =
 \frac{\bar{\Omega}^{2}\xi^{4-s}}{q_{0}}
 \left[
 -b^{4}_{1} q_{1} + o(\xi^{2 s}) \right] = 0
 \end{aligned}
\end{equation}
where $q_{0} = 4b^{3}_{1}s^{3}(s+1)^{3}$ and $q_{1} = (s+15)(s-2)$. Now the only acceptable solution is $s = 2$, and the functions near a Type II point behave as follows (see \autoref{fig:App-BehaviourSingularities})
\begin{equation}
 \label{eq1:SubstitutionRespFunS2}
  \begin{split}
       v_{\ast}\left( \xi \right) & =   b_{0}\xi^{-2} \: , \quad
   v_{\ast, x}\left( \xi \right)   = - b_{1}\xi^{-3} \: , \quad
  v_{\ast, xx}\left( \xi \right)   =   b_{2}\xi^{-4} \\
       y_{\ast}\left( \xi \right) & =   b_{3}\xi^{2}  \: , \quad
       M_{\ast}\left( \xi \right)   =   b_{4}\xi^{4}  \: , \quad
       T_{\ast}\left( \xi \right)   =   b_{5}\xi^{3}
  \end{split}
\end{equation}
where the expression of the $b_{i}$ coefficients can be found in \cite{olhoff_76a_optimizationVibrationsHighOrderFrequencies}.


\begin{small}
 \bibliographystyle{spmpsci} 
 \bibliography{biblioMendeley}
\end{small}
\end{document}